\documentclass[oldversion]{aa}
\usepackage{txfonts}
\usepackage{graphicx}
\usepackage{natbib}
\usepackage{rotating}
\bibpunct{(}{)}{;}{a}{}{,}

\newcommand{\rpms}[1]{{#1}\,\mathrm{rad\,m}^{-2}}

\begin{document}

\title{The Westerbork SINGS survey III.}
\subtitle{Global magnetic field topology}

\author{R. Braun\inst{1}
   \and G. Heald\inst{2}
   \and R. Beck\inst{3}}

\offprints{Robert Braun, \email{robert.braun@csiro.au}}

\institute{CSIRO -- Astronomy and Space Science, PO Box 76, Epping, NSW 1710, Australia 
      \and Netherlands Institute for Radio Astronomy (ASTRON), Postbus 2, 7990 AA Dwingeloo, The Netherlands
      \and Max-Planck-Institut f\"ur Radioastronomie, Auf dem H\"ugel 69, 53121 Bonn, Germany}

\date{Received $nnn$ / Accepted $nnn$}

\abstract{ A sample of large northern Spitzer Infrared Nearby Galaxies
  Survey (SINGS) galaxies was observed with the
  Westerbork Synthesis Radio Telescope (WSRT) at 1300 -- 1760 MHz. In
  Paper II of this series, we described sensitive observations of the
  linearly polarized radio continuum emission in this WSRT-SINGS
  galaxy sample. Large-scale magnetic field structures of two basic
  types are found: (a) disk fields with a spiral topology in all
  detected targets; and (b) circumnuclear, bipolar outflow fields in
  a subset. Here we explore the systematic patterns of azimuthal
  modulation of both the Faraday depth and the polarized intensity and
  their variation with galaxy inclination. A self-consistent and fully
  general model for both the locations of net polarized emissivity at
  1 -- 2 GHz frequencies and the global magnetic field topology of nearby
  galaxies emerges. Net polarized emissivity is concentrated into two
  zones located above and below the galaxy mid-plane, with the
  back-side zone suffering substantial depolarization (by a factor of
  4 -- 5) relative to the front-side zone in its propagation through
  the turbulent mid-plane. The field topology which characterizes the
  thick-disk emission zone, is in all cases an axisymmetric spiral
  with a quadrupole dependance on height above the mid-plane. The
  front-side emission is affected by only mild dispersion (10's of
  rad~m$^{-2}$) from the thermal plasma in the galaxy halo, while the
  back-side emission is affected by additional strong dispersion (100's of
  rad~m$^{-2}$) from an axi-symmetric spiral field in the galaxy
  mid-plane. The field topology in the upper halo of galaxies is a mixture
  of two distinct types: a simple extension of the axisymmetric
  spiral quadrupole field of the thick disk and a radially
  directed dipole field. The dipole component might be a manifestation
  of (1) a circumnuclear, bipolar outflow, (2) an {\it in situ}
  generated dipole field, or (3) evidence of a non-stationary global
  halo.  }

\maketitle

\section{Introduction\label{section:intro}}

The magnetic fields in spiral galaxies are an important component, but
their basic three dimensional topology remains largely unknown. Two
of their main characteristics are however, known. First, the fields in relatively
face-on spiral galaxies are seen to follow the spiral pattern traced
in the optical morphology. In the handful of more edge-on galaxies
that have been imaged to date, the field distributions are seen to
extend into the halo regions, and have a characteristic {\sf X}-shaped
morphology \citep[eg.][]{heesen_etal_2009}. Apart from these 
basic properties, the details of the magnetic field topology are unknown.

Observations of polarized flux, polarization vector orientations, and
Faraday rotation measures all provide information about the magnetic
field associated with different electron populations and at
different projections with respect to the line of sight. Synchrotron
emission originates in ultrarelativistic electrons spiralling around
magnetic field lines, is beamed in the direction of motion of the
electron, and is polarized perpendicular to the orientation of the
field line. Polarized synchrotron radiation and polarization vector
orientation are thus direct tracers of the magnetic fields
perpendicular to the line-of-sight (LOS), $B_{\perp}$, within the
region where both ordered magnetic fields and relativistic electrons
are maximized. The Faraday rotation measure (RM), or more generally
the Faraday depth, $\Phi$, that pertains to a given component of
polarized emission, is sensitive to the integrated product of magnetic field
component parallel to the LOS ($B_{\parallel}$) and the thermal
electron density in the foreground of a polarized emission component:
\begin{equation}
\Phi\,\propto\,\int_{\mathrm{source}}^{\mathrm{telescope}}n_e\vec{B}{\cdot}d\vec{l}.
\end{equation}
The Faraday depth is defined to be positive when $\vec{B}$ points
toward the observer, and negative when $\vec{B}$ points away. When
assessing the magnetic field geometry traced by these observational
characteristics, it is essential to keep in mind that the observable
attributes may originate in distinct regions of space. The classical
Faraday rotation measure (RM) is an observable quantity derived from
the polarization angle difference(s) $\Delta\chi$ between two (or
more) frequency bands as
$RM=\Delta\chi/(\lambda_1^2-\lambda_2^2)$. The empirically determined
RM is only equivalent to the Faraday depth $\Phi$ for a simple
background emitter plus foreground dispersive screen geometry.

Polarized emission can become depolarized in a number of ways: beam
depolarization can arise because the spatial resolution element is
large relative to the size of significant variations in the field
orientation or the thermal electron content, while Faraday
depolarization can arise because synchrotron emission and Faraday
rotation take place in the same extended volume along the
LOS. Polarized emission from different locations (either separated
spatially or along the LOS) is affected by different amounts of Faraday
rotation, such that at a given wavelength there may be orthogonal
polarization angles that cancel, yielding no net polarization at that
wavelength. Beam depolarization can be circumvented in principle by
using higher angular resolution, although the brightness sensitivity
may then be insufficient to detect the extended emission at
all. Faraday depolarization can be circumvented in principle by
achieving a sufficiently complete sampling of the $\lambda^2$
measurement space (relevant for measurements of RM and $\Phi$), since
cancellation effects are confined to discrete wavelengths or ranges of
wavelength.


All of these observables can be used to constrain the likely magnetic
field topology in the galaxies observed. In a previous paper
(\citet{heald_etal_2009}, hereafter Paper II), we presented our
polarimetric results for a large sample of nearby galaxies observed to
a comparable sensitivity limit of about 10~$\mu$Jy beam$^{-1}$ RMS. In
this paper, we begin by briefly summarizing the observations and data
reduction steps of Paper II in Sect.~\ref{section:reductions}. Trends
noted in the data are described in Sect.~\ref{section:trends}.  In
Sect.~\ref{section:Bdist}, we then explore how particular magnetic field
geometries might relate to the observations. We conclude
the paper in Sect.~\ref{section:disc}.

\section{Summary of observations and data reduction\label{section:reductions}}

The observational parameters and data reduction techniques of the
WSRT-SINGS survey were presented in detail both by
\citet{braun_etal_2007}, and specifically regarding the polarization
data in Paper II. Here we recap the most important details. For
more information, the reader is referred to Sect.~2 and Appendix~A of
Paper II.

The data used in this analysis were obtained using the Westerbork Synthesis
Radio Telescope (WSRT). Two observing bands were used: of $1300-1432$ MHz
and $1631-1763$ MHz (centered on 22- and 18-cm, respectively), there being 512 channels in each
band and in all four polarization products. Each galaxy
in the WSRT-SINGS sample (refer to Paper II) was observed for 12~hr in the 
Maxi-short configuration of the WSRT. During each 12~hr synthesis, the observing
frequency was switched between the two bands every 5~min. This provided an
effective observation time of 6~hr per band, and good $uv$
coverage in both bands.

The data for both bands were analyzed using the Rotation Measure
Synthesis (RM-Synthesis) technique
\citep[][see also Paper II]{brentjens_debruyn_2005}. This provides the 
possible reconstruction of the intrinsic polarization vectors along each LOS,
within the constraints set by the observing frequencies. The output of the
RM-Synthesis procedure was deconvolved along the Faraday depth ($\Phi$) axis,
as described in Paper II. Polarized fluxes, polarization angles, and Faraday
depths were extracted from these data and are discussed for
each target galaxy in Paper II. In that paper, we also estimate the
contribution to the RM from the Milky Way foreground using only
background radio sources in the observed fields, rather than the target
galaxies themselves. With this collection
of data, we noted several patterns in the target galaxies, and that the
basic patterns were common to the sample galaxies collectively. In this
paper, we seek to explain these patterns using a common global magnetic
field topology.

\section{Observational trends\label{section:trends}}

Several interesting patterns emerge from our study of a large sample
of galaxy types. Polarized emission is found to originate both in the
disks of actively star-forming galaxies and in what appear to
be either AGN or star-formation driven (circum-)nuclear or galactic
wind outflows. ``Disk'' emission is relatively planar and
detected out to large radii, whereas apparent ``outflow'' components extend
much further from the plane and are only detectable at small radii. The
magnetic field orientations are in all cases simply related to the
morphology of these components (as shown in Fig. 4 of Paper II).

Disk fields have a spiral morphology that is strongly correlated with
the orientation and pitch angle of traditional tracers of spiral arms,
such as massive stars and dust lanes. This is despite the
polarized emission possibly being both coincident with,
and independent of, other spiral arm tracers. Classic examples of
disk-dominated fields can be seen in galaxies such as NGC 628, 5194, and
6946, although they are present in almost all of our targets with
varying detectability. On the other hand, outflow-related fields are
typically oriented along the periphery of the bipolar lobes, and are often
brightest close to the disk. The contribution of outflow-related
components may be apparent in NGC 4569, 4631, and possibly 4736.

Here, we illustrate and describe some of the trends present in the
WSRT-SINGS dataset. We consider the polarized flux distribution
(Sect.~\ref{subsection:polaz}), the Faraday depth distribution
(Sect.~\ref{subsection:fdaz}), the effects of depolarization in
(Sect.~\ref{subsection:depol} and Sect.~\ref{subsection:2disk}). In
Sect.~\ref{section:Bdist}, we explore the predictions of a variety of
global magnetic field topologies that
can be used to model these trends.

\subsection{Polarized flux\label{subsection:polaz}}

\begin{figure*}
\resizebox{\hsize}{!}{\includegraphics{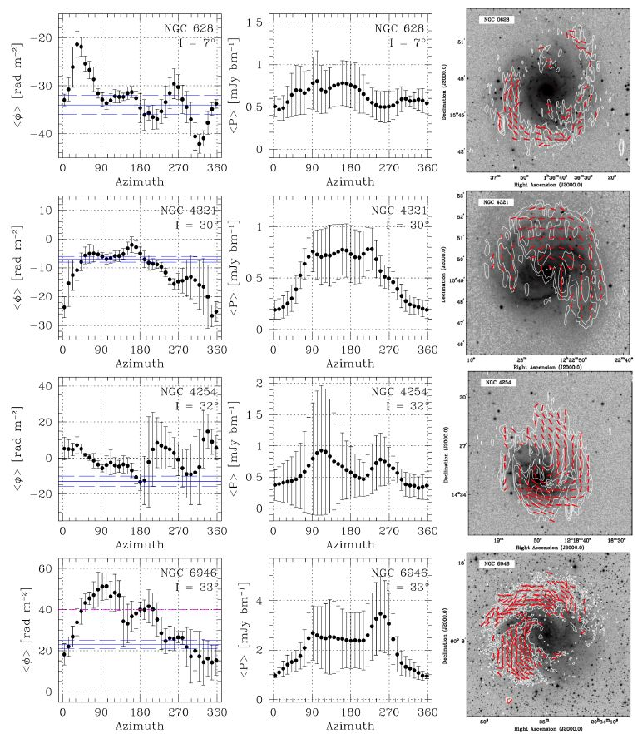}}

\caption{Azimuthal variation in peak polarized intensity (center
  panel) and associated Faraday depth (left panel) for galaxies with
  extended polarized emission (illustrated in the right panel from
  Fig.~5 of Paper II). The mean values in azimuthal wedges, each
  subtending 10$^\circ$ within the galaxy disk, are plotted with error
  bars giving the wedge RMS. Galaxies are arranged in order of
  increasing inclination (top to bottom) from face-on. Azimuth is
  measured counter-clockwise from the receding major axis.}
\label{figure:fdpofaz}
\end{figure*}

\addtocounter{figure}{-1}
\begin{figure*}
\resizebox{\hsize}{!}{\includegraphics{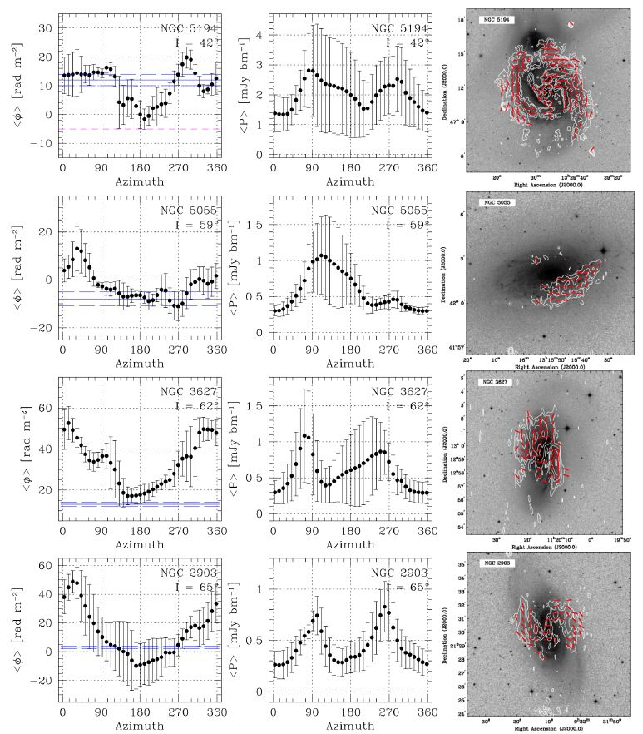}}

\caption{(continued) Azimuthal variation in peak polarized intensity (center
  panel) and associated Faraday depth (left panel) for galaxies with
  extended polarized emission (illustrated in right panel).}
\end{figure*}

\addtocounter{figure}{-1}
\begin{figure*}
\resizebox{\hsize}{!}{\includegraphics{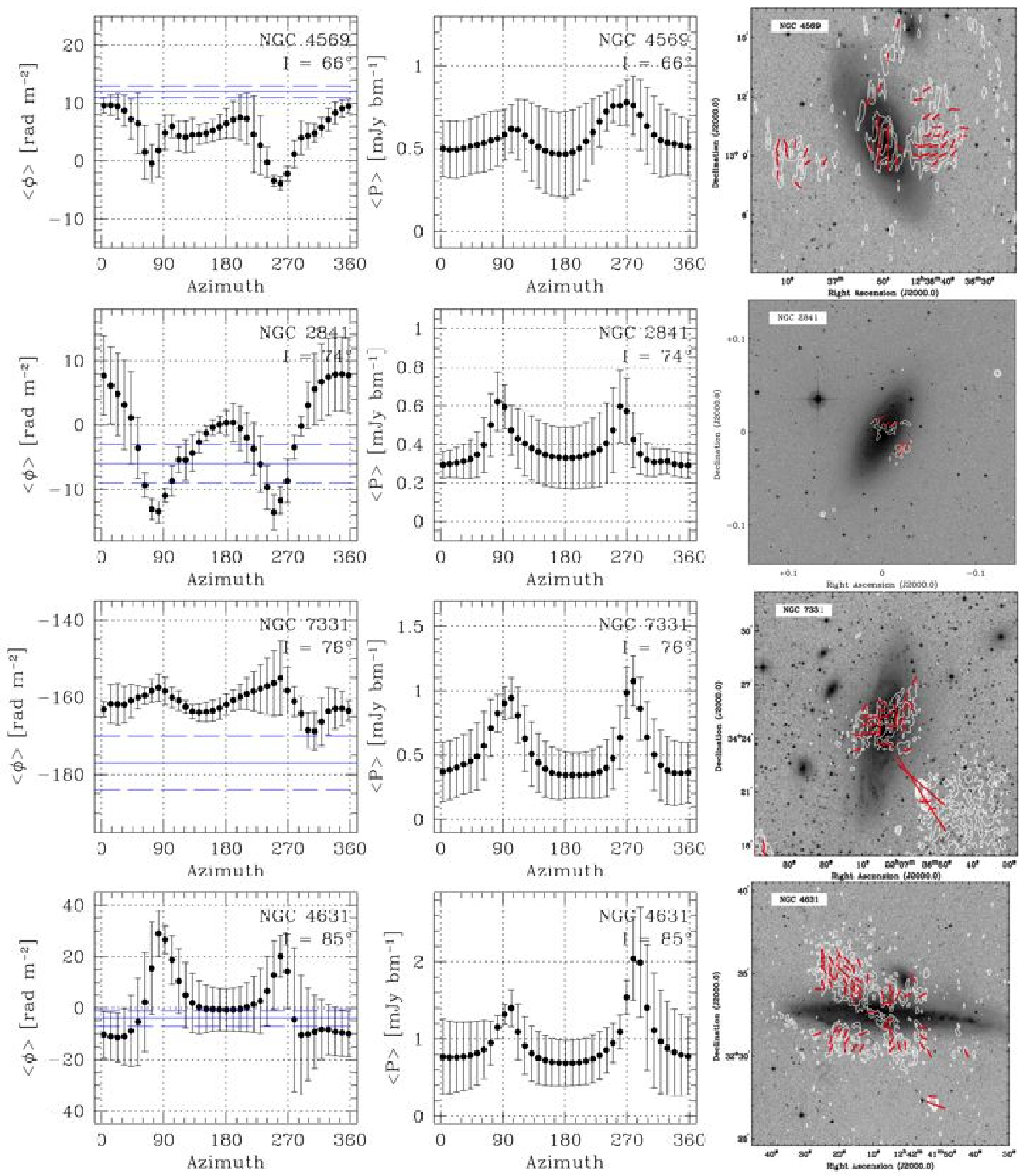}}

\caption{(continued) Azimuthal variation in peak polarized intensity (center
  panel) and associated Faraday depth (left panel) for galaxies with
  extended polarized emission (illustrated in right panel).} 
\end{figure*}

A remarkable pattern emerges for our sample relating to the basic
distribution of polarized intensity in galaxy disks at GHz
frequencies. For small inclination angles, there is a general
gradient in the average polarized intensity that is approximately
aligned with the major axis of the target galaxy. This
gradient from high to low polarized intensity has the same sign in all
well-detected cases, from high values on the kinematically
approaching major axis to low values on the receding major axis. This
effect cannot be explained by a symmetric planar field
geometry. As the inclination of the target galaxy increases,
a pair of local maxima in polarized intensity begins to separate from
the approaching major axis and propagates toward the minor axis. This
is simply a geometrical effect because maximum polarized emission
emerges from magnetic fields perpendicular to the line of sight, which,
in the case of a planar field geometry, are strongest near the minor
axis \citep{stil_etal_2009}. Even when the inclination has become
quite substantial, and the two local maxima of polarized intensity are
near the minor axis, there is still a systematic tendency for the
receding major axis to have the overall minimum polarized brightness.

This pattern in the distribution of polarized flux in our target
galaxies is visible first of all in Fig.~4 of Paper II, which shows
maps of the polarized flux in each galaxy. We quantify this
trend in Fig.~\ref{figure:fdpofaz} by plotting the average peak
polarized intensity, $<P>$, and the associated Faraday depth,
$<\Phi>$, within inclination-corrected wedges spanning 10 degrees of
azimuth in the disk of each galaxy in which extended polarized
emission was detected. Azimuth is measured counter-clockwise from the
kinematically receding major axis (see Table~1 of Paper~II for basic
data on each target). Error bars on the points in the plots represent
the RMS variation within each wedge. The estimated foreground
contribution to $<\Phi>$ and its error caused by our own Galaxy toward
each target is indicated by the horizontal lines. The pairs of panels
in the figure have been ordered by increasing galaxy inclination (top
to bottom) extending from less than 10 degrees for NGC~628 to about 85
degrees in NGC~4631.

The simple trend described above is clear, the polarized intensity
showing a global minimum toward the receding major axis and there
being a
systematic progression from one broad maximum at the approaching major
axis to a pair of maxima that move to the minor axes as the
inclination increases. The latter effect is due to geometry of a
planar field, while the former one has no current explanation. We note that
the polarization asymmetry along the major axis is not as pronounced
at frequencies of 5~GHz and higher (see e.g.  \citet{beck_2007}).

\subsection{Faraday depth\label{subsection:fdaz}}

The variation in Faraday depth with azimuth (shown in
Fig.~\ref{figure:fdpofaz}) also shows systematic trends with
increasing inclination, although not as cleanly as those seen in
polarized intensity. One complication is that each Faraday
depth distribution can be either positive or negative, so both options
need to be considered in assessing a possible trend with
inclination. Another complication is the uncertainty in the
foreground contribution to $\Phi$, which is critical in distinguishing
peaks (be they positive or negative) in the modulation patterns from
minima (consistency with that foreground level). What is immediately
clear from an assessment of the $<\Phi(\phi)>$ patterns is that they
are in no case consistent with being symmetric sinusoids (either of
period $2\pi$ or $\pi$) in their excursions about the estimated
foreground Faraday depth. The implication is that a thin axisymmetric
or bisymmetric spiral disk is not a viable model for the medium
responsible for the Faraday rotation of the polarized emission
detected at these frequencies \citep[cf.][]{krause_1990}.

The $<\Phi(\phi)>$ pattern that applies to a large fraction of our
sample has a minimum Faraday depth (consistent with the estimated
foregound) that occurs close to the approaching major axis (at an azimuth of
180$^\circ$) and a single maximum excursion near the receding major
axis at low inclinations. This same pattern applies to many of the
eight lowest inclination galaxies and would also apply to NGC~6946 if
the previously published estimate of the Galactic foreground value,
$\Phi_{FG} = \rpms{40}$ by \citet{beck_2007}, were the correct one,
rather than the $\Phi_{FG} = \rpms{23}\pm2$ we estimate in Paper~II. A
similar consideration applies to NGC~5194, for which
\citet{horellou_etal_1992} estimated $\Phi_{FG} = \rpms{-5}\pm12$
rather than our estimate of $\Phi_{FG} = \rpms{+12}\pm2$. We 
plotted these alternate estimates of $\Phi_{FG}$ in the figure with
horizontal lines. We also note here a typographical error in the
value of $\Phi_{FG}$ for NGC~4321 in Table~1 of Paper~II, which is
$\rpms{-7}$ and not $\rpms{-17}$ as stated there. This
pattern of maximum and minimum is not a simple sinusoid but has a
clear asymmetry about an azimuth of 180$^\circ$ that is most obvious
in NGC~4321 and 6946. When the inclination exceeds 65$^\circ$, there
is a sudden change to a pattern of two peaks near the minor axis that
is found in all four of the highly inclined galaxies in our sample. It
may well be of further significance that several (and possibly all) of
these highly inclined galaxies had already been identified as having a
morphology suggestive of a circum-nuclear or galactic wind outflow.

Another critical observation is the magnitude of the Faraday depth
excursions from the foreground value, which is in all cases very
modest, typically between 10 and 30~rad~m$^{-2}$. This is
substantially less than the Faraday depth variations measured through
the entire disk of the Large Magellanic Cloud by
\citet{gaensler_etal_2005} using distant background sources that show
average excursions of plus and minus 50~rad~m$^{-2}$ and peak
excursions of $+$245 and $-$215~rad~m$^{-2}$. Since the LMC is in no
way a remarkable galaxy in terms of its likely field strength (based
on the synchrotron surface brightness) or thermal electron populations
(based on the star-formation-rate density) relative to our sample
galaxies, the implication is that the medium responsible for the
Faraday rotation of the diffuse polarized emission in our targets only
extends over a small fraction of the complete line-of-sight. A similar
conclusion was reached by \citet{berkhuijsen_etal_1997} for NGC~5194
based on the smaller Faraday depths seen at 1.4-1.8 GHz relative to
5-10~GHz for that target.

An important conclusion that follows directly from a comparison
of the $<\Phi(\phi)>$ and $<P(\phi)>$ plots is that the polarized
intensity and the Faraday rotation in the highly inclined galaxies of
our sample must originate in distinct regions along the
line-of-sight. This is because $\Phi$ is proportional to B$_\parallel$,
while $P$ is proportional to $B^{1+\alpha}_{\perp}$. For any given
field geometry, a peak in B$_\parallel$ will be accompanied by a
minimum in $B^{1+\alpha}_{\perp}$ and vice versa. The observation that
maximal excursions of both $<\Phi>$ and $<P>$ are seen in all four of
the most inclined galaxies cannot be achieved with any co-extensive
geometry of the emitting and dispersing media.

\subsection{Depolarization\label{subsection:depol}}

In addition to detecting a systematic pattern of Faraday depth
excursions across the full LMC disk, \citet{gaensler_etal_2005} also
document the very substantial depolarizing effect of the LMC disk on
background polarized sources observed at 1.4~GHz. Despite a likely
mean angular source size of only about 6 arcsec, which projects to
1.5~pc at the LMC disk, these sources are more depolarized
(by a factor of more than two) when the LOS is associated with thermal
electrons in the LMC disk exceeding an emission measure of about
50~pc~cm$^{-6}$. The implication appears to be that significant RM
fluctuations are present at the 1.4~GHz observing frequency on scales
$<<$~1.5~pc. Since even the diffuse ionized gas that permeates galaxy
disks has an emission measure in excess of about 20~pc~cm$^{-6}$ in
moderately face-on systems \citep[e.g.,][]{greenawalt_etal_1998}, we can
expect a significant degree of depolarization for any ``backside''
emission in our galaxy sample. This will be true in particular for the
diffuse polarized emission originating in the target galaxy itself,
since the relevant scale is then that of the observing beam,
$\ge$15~arcsec, which projects across more than 700~pc at the typical
galaxy distance of 10~Mpc.

It therefore seems likely that the polarized intensity from low
inclination galaxy disks observed at GHz frequencies is dominated by
emission from that portion of the galaxy disk/halo that faces
us. Corresponding structures from the far side of the galaxy would be
dispersed and depolarized by turbulent magneto-ionic
structures in the star-forming mid-plane. We recall that it is only in
those regions dominated by a regular rather than a turbulent magnetic
field that a net polarized emissivity is expected at all. Given
the strong concentration of massive-star formation and its associated
turbulent energy injection into the mid-plane, there may well be two
zones of enhanced net polarized emissivity that are offset above and
below the mid-plane. Each region of polarized emissivity will then
experience the dispersive effects of its own line-of-sight
foreground, which is likely to be dominated by {\it thermal\ } rather
than {\it relativistic\ } plasma. For the near-side polarized emission,
this is likely to be caused by the extended thermal halo of the host
galaxy, while for the far-side polarized emission there will be
the additional contribution of the dense (depolarizing and dispersive)
mid-plane.

\subsection{The ``second'' polarized disk\label{subsection:2disk}}

\begin{figure*}
\resizebox{\hsize}{!}{\includegraphics{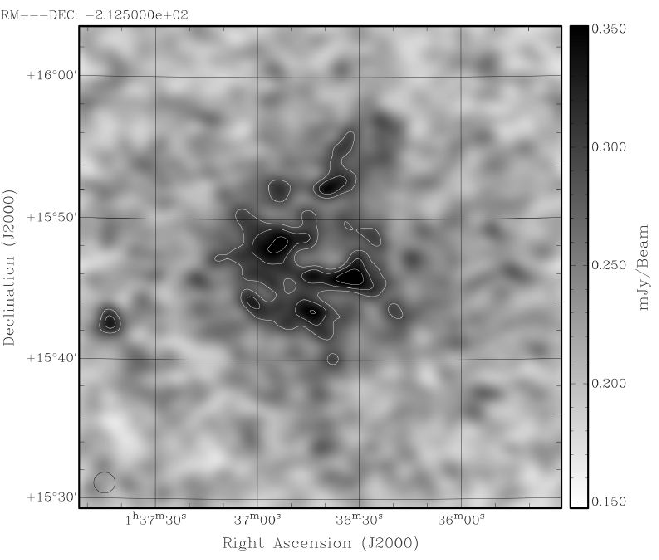},\includegraphics{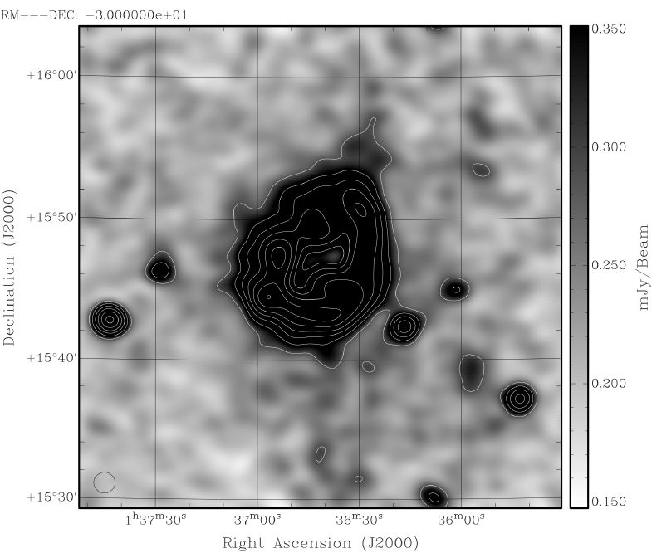},\includegraphics{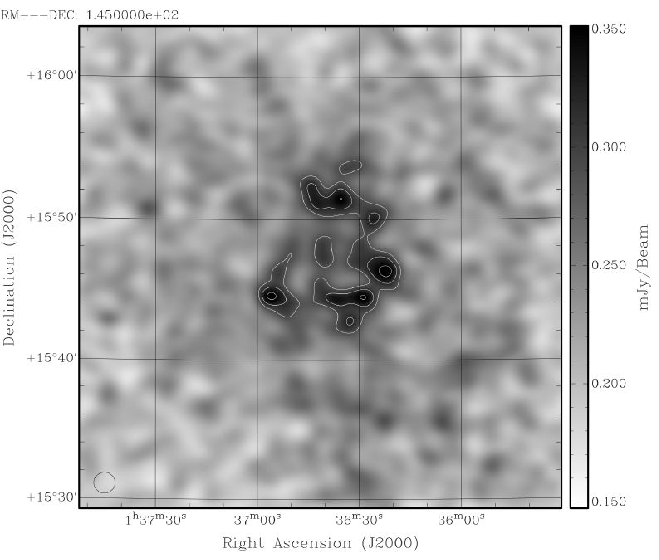}}
\caption{Polarized intensity at distinct Faraday depths toward NGC
  628. The dominant Faraday depth component, centered near $\rpms{-30}$
  is shown in the center panel, while the two secondary components
  centered near $-$213 and $\rpms{+145}$ are shown on the left and
  right. The greyscale varies as indicated. The contours begin at
  0.29~mJy~beam$^{-1}$ and increase by factors of 1.1 for the
  secondary components and 1.3 for the primary component.}
\label{figure:n628smo}
\end{figure*}

\begin{figure*}
\resizebox{\hsize}{!}{\includegraphics{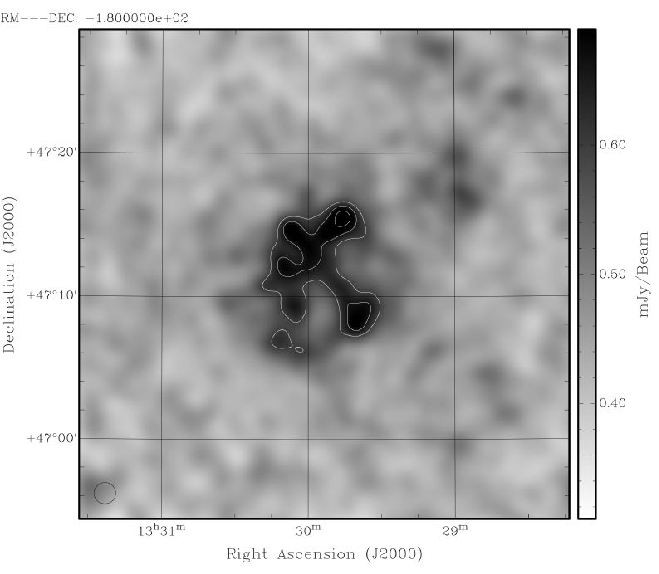},\includegraphics{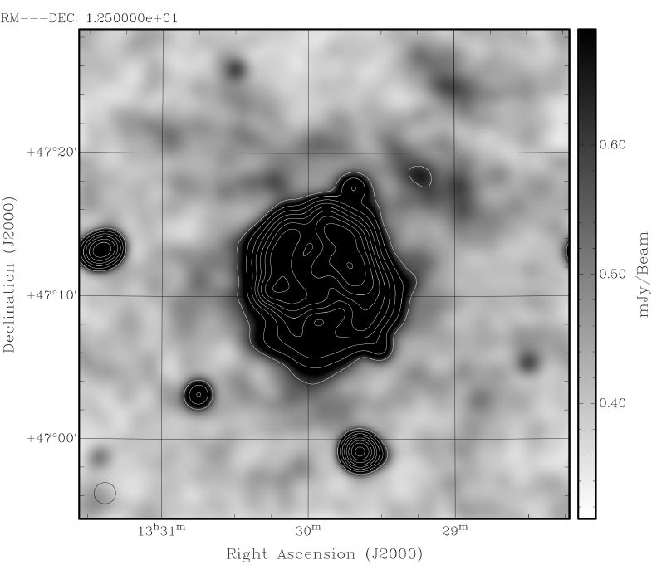},\includegraphics{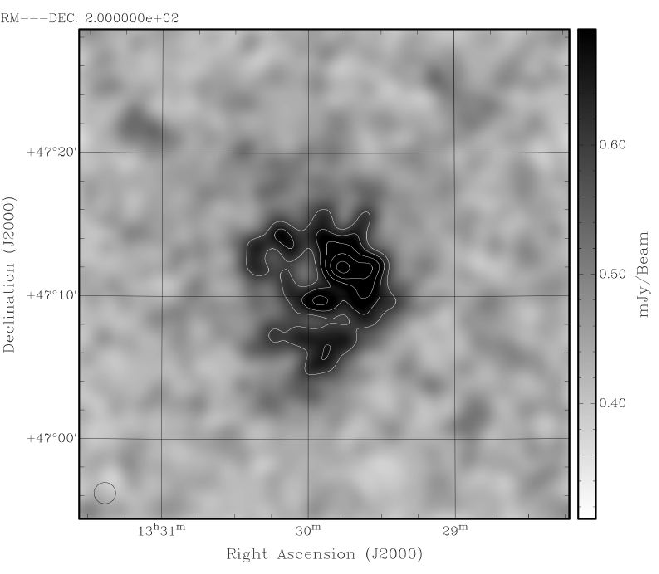}}
\caption{Polarized intensity at distinct Faraday depths toward NGC
  5194. The dominant Faraday depth component, centered near $\rpms{+13}$
  is shown in the center panel, while the two secondary components
  centered near $-$180 and $\rpms{+200}$ are shown on the left and
  right. The greyscale varies as indicated. The contours begin at
  0.6~mJy~beam$^{-1}$ and increase by factors of 1.1 for the secondary
  components and 1.3 for the primary component.}
\label{figure:n5194smo}
\end{figure*}

\begin{figure*}
\resizebox{\hsize}{!}{\includegraphics{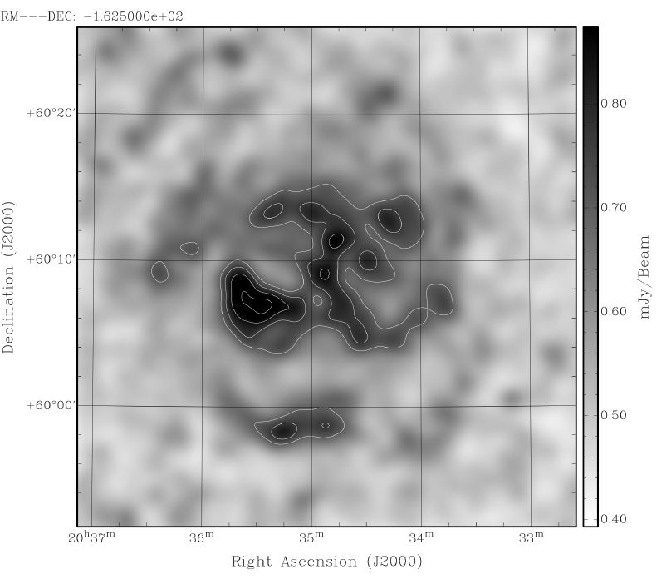},\includegraphics{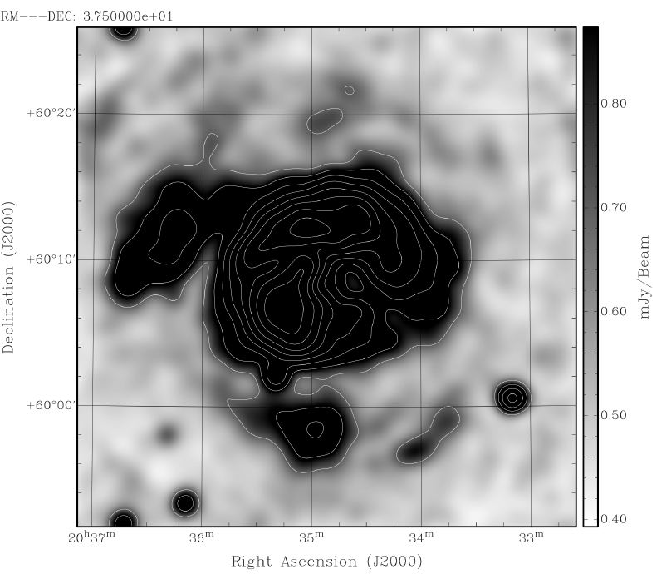},\includegraphics{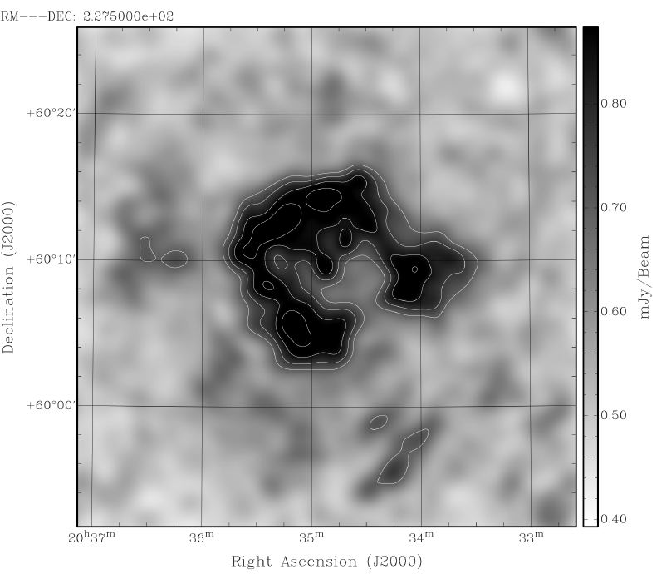}}
\caption{Polarized intensity at distinct Faraday depths toward NGC
  6946. The dominant Faraday depth component, centered near $\rpms{+38}$
  is shown in the center panel, while the two secondary components
  centered near $-$162 and $\rpms{+228}$ are shown on the left and
  right. The greyscale varies as indicated. The contours begin at
  0.7~mJy~beam$^{-1}$ and increase by factors of 1.1 for the secondary
  components and 1.3 for the primary component.}
\label{figure:n6946smo}
\end{figure*}

Because of the likely Faraday depth of the mid-plane medium, it is
conceivable that the near- and far-side polarized emission zones in
moderately face-on galaxies would experience very different Faraday
dispersion, making it possible to distinguish the two components along
each line-of-sight usingthe RM synthesis technique, which we employed
in our study. We conducted a deep search for multiple Faraday depth
components along the line-of-sight to detect disk emission from each
of our sample galaxies, after applying a spatial smoothing to the
$P(\phi)$ cubes that results in an angular beamsize of 90
arcsec. Solid detections of faint secondary emission components were
made in the brightest face-on galaxies of our sample, as shown in
Figs.~\ref{figure:n628smo}--\ref{figure:n6946smo}. In addition to the
bright polarized emission that typically resides at a Faraday depth
within only a few tens of rad~m$^{-2}$ of the Galactic foreground
value, much fainter polarized emission (by a factor of 4 -- 5) is
detected in NGC~628, 5194, and 6946, which is offset to both positive
and negative Faraday depths by about 200 rad~m$^{-2}$. We first
considered whether these faint secondary components might be
instrumental in nature, since they are similar to the Faraday depth
side-lobes of our instrumental response (as described in Paper II),
but concluded that they are likely reliable.  Indicators of the
reliability of these faint components are that (1) they do not occur
toward the brightest low dispersion components; (2) the Faraday
depth separation of the secondary components varies from source to
source, while the instrumental sidelobe response does not; and (3)
the faint positive and negative-shifted components form a
complementary distribution of eachother, rather than merely
repeating themselves in detail.

Detection of this highly dispersed and probably depolarized ``second
disk'' supports the emerging model in which the polarized
intensity observed at GHz frequencies from nearby galaxy disks is
dominated by a region of emissivity located above the mid-plane on the
near-side, which subsequently experiences Faraday rotation within the
extended halo of the galaxy amounting to only a few tens of
rad~m$^{-2}$. Significantly fainter polarized emission (by a factor of
4 -- 5) is detected from the farside of the mid-plane, which displays
the additional dispersive effects of that mid-plane zone amounting to
plus and minus 150 -- 200 rad~m$^{-2}$ within the three galaxies where
this could be detected.

\subsection{Small-scale RM fluctuations\label{subsection:RMfluc}}

\begin{figure*}
\resizebox{\hsize}{!}{\includegraphics{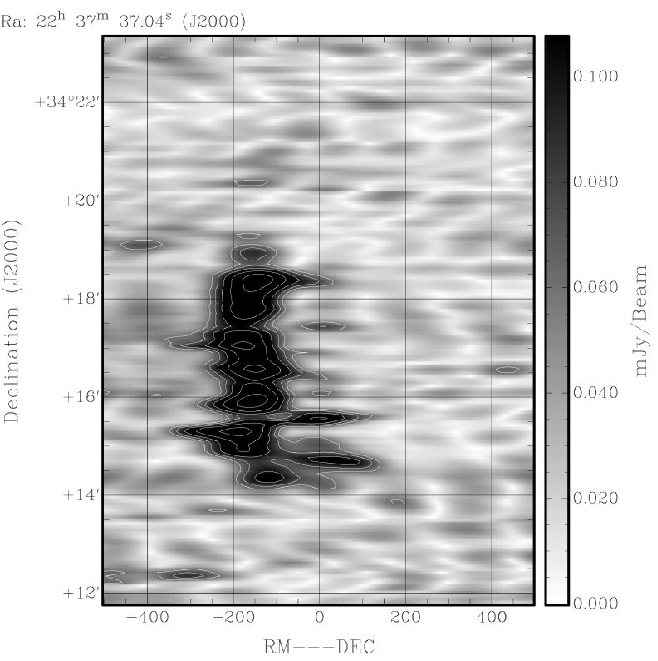},\includegraphics{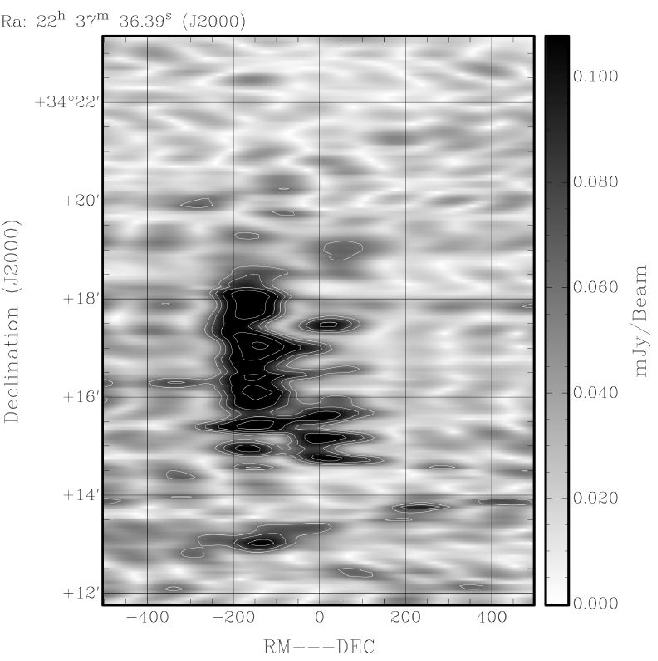},\includegraphics{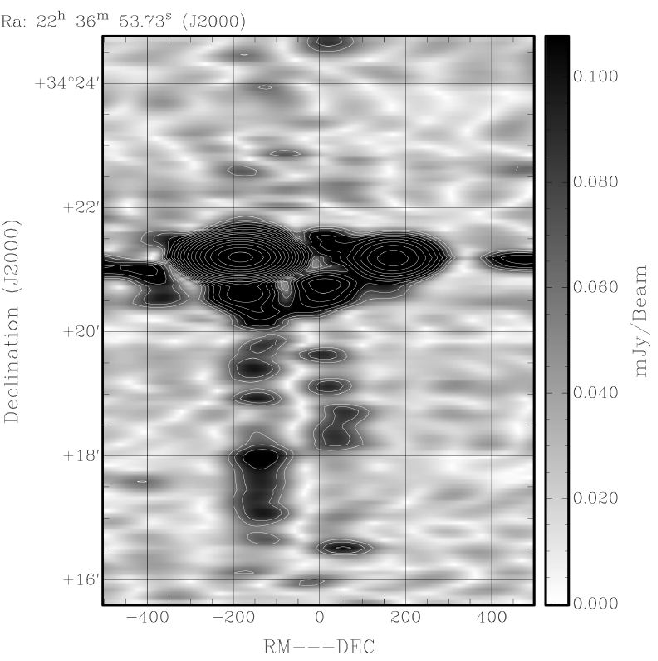}}
\caption{Polarized intensity as a function of Faraday depth and
  declination in the field of NGC 7331. The two left hand panels are
  at right ascensions that intersect the diffuse lobe of a background
  head-tail radio galaxy. The right-hand panel intersects a background
  double-lobed radio galaxy as well as a region of diffuse Galactic
  polarized emission.}
\label{figure:n7331FDD1}
\end{figure*}

\begin{figure*}
\resizebox{\hsize}{!}{\includegraphics{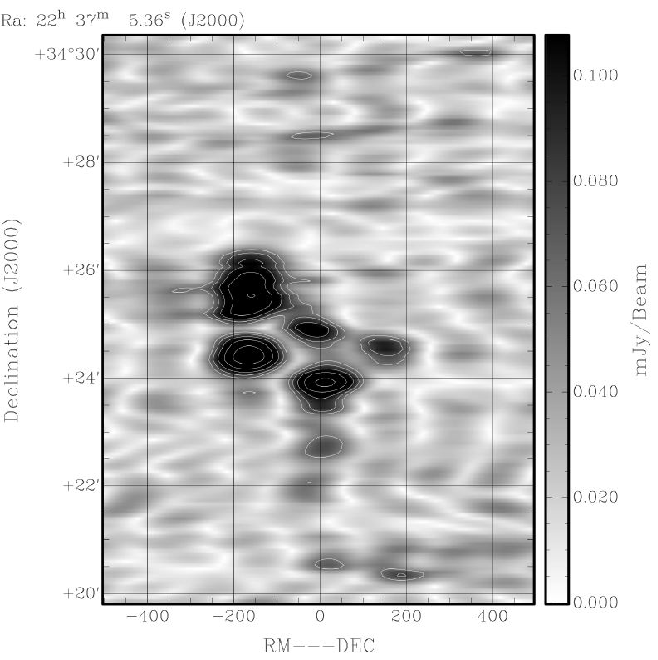},\includegraphics{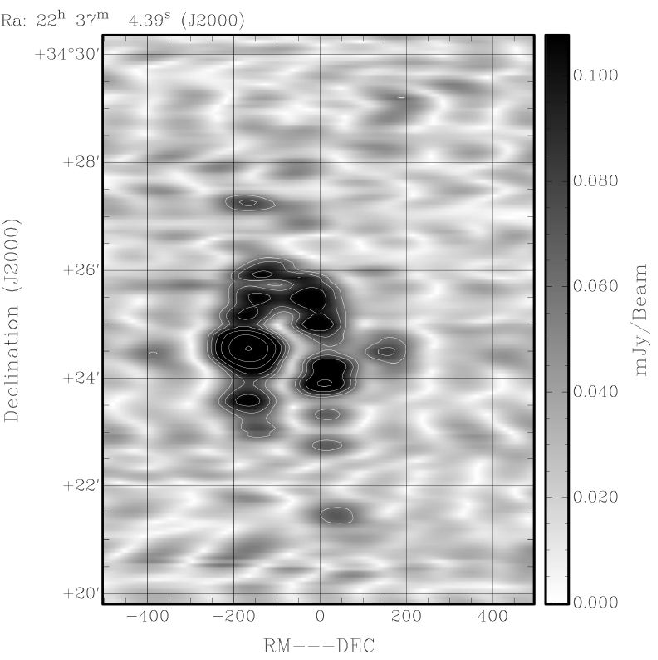},\includegraphics{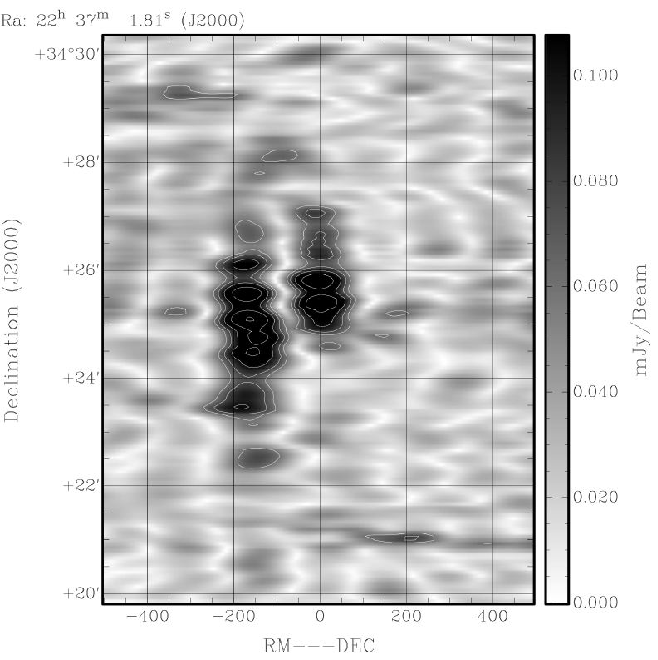}}
\caption{Polarized intensity as a function of Faraday depth and
  declination in the field of NGC 7331. Slices are presented at three
  right ascensions that intersect the disk of NGC~7331 near the major
  axis.}
\label{figure:n7331FDD2}
\end{figure*}

Extremely small-scale RM fluctuations within our own
Galaxy were discovered in our sample data for the field of
NGC~7331. This target is observed through the Galaxy near $(l,
b)~\sim~(94, -21)$, within an extended ($30^\circ\times30^\circ$)
region of particularly large negative RMs near $\rpms{-200}$
\citep[e.g.,][]{johnston-hollitt_etal_2004}. A corresponding region of
large positive Galactic RMs is centered near $(l, b)~\sim~(250,
-10)$. These two regions correspond to the directions where we look
directly along what is likely to be the axisymmetric spiral field of the
Galaxy \citep[e.g.,][]{sun_etal_2008}. Diffuse polarized emission from
the lobes of a background head-tail radio galaxy, the disk of NGC~7331,
and even the Galactic synchrotron itself display the remarkable
oscillatory behavior of the Faraday depth of the polarized emission
with position in the field. As shown in the Faraday depth versus
declination slices of Figs.~\ref{figure:n7331FDD1} and
\ref{figure:n7331FDD2}, there are two dominant Faraday depths present
in this field, one near $\rpms{-180}$ and the other near
$\rpms{0}$. Depending on the exact location along the indicated
declination slice, either one or the other of these RMs are
encountered. In some regions, the transition from one RM to the other
is well-resolved and a single RM value is observed over several
beam-widths (of about 20 arcsec or 0.1~pc at a distance of 1~kpc),
while in other regions the transition is unresolved, such that both
RMs overlap spatially with only the peak polarization showing the
oscillation between the two values. When completely unresolved
extra-galactic sources are observed in this field, they display only
one or the other of these two possible foreground RMs (compare Table 2
of Paper II), but more extended sources show an oscillation between
the two values, $\rpms{-180}$ occurring 3 -- 4 times as often as
$\rpms{0}$. The regular Galactic magnetic field appears to be
organized into discrete filamentary components with transverse sizes
significantly less than a parsec. 

\section{Magnetic field distributions\label{section:Bdist}}

The observational results summarized in Sect.\,\ref{section:trends} suggest
a magnetic field geometry dominated by an in-plane
field described by an axisymmetric spiral structure (ASS) and/or possibly
a bi-symmetric spiral structure (BSS). These topologies assume a planar
field configuration (in the $(x,y)$ plane) given by a family of
logarithmic spirals defined in radial coordinates $(r,\phi)$
by the form
\begin{equation}
r = a e^{b(\phi+c)},
\end{equation}
for a radial scaling constant, $a$, an angular scaling
constant, $b$, which is related to the spiral pitch angle, $\psi^\prime_{xy}$, by,
\begin{equation}
\psi^\prime_{xy} = \tan^{-1}(b)
\end{equation}
and an angular offset, $c$, which defines each curve in the family. We
find it convenient to use the complement of the spiral pitch
angle, $\psi_{xy}$ given by,
\begin{equation}
\psi_{xy} = 90^\circ - \psi^\prime_{xy} = \tan^{-1}(1/b).
\label{eqn:pxy}
\end{equation}
The right-handed cartesian components of an inward directed planar ASS
logarithmic spiral are then given by,
\begin{eqnarray}
B_x & = & B\cos(\phi+\psi_{xy}) \\
B_y & = & B\sin(\phi+\psi_{xy}) \\
B_z & = & 0.
\end{eqnarray}
For the corresponding case of a planar BSS magnetic field with the
usually assumed sinusoidal modulation of $B$ with $\phi$, we have,
\begin{eqnarray}
B_x & = & B\cos(2\phi+\psi_{xy}+\mu) \\
B_y & = & B\sin(2\phi+\psi_{xy}+\mu) \\
B_z & = & 0, 
\end{eqnarray}
where $\mu$ is used to track the positive peak in the field modulation
pattern from an initial value $\mu_0$ such that,
\begin{equation}
\mu = \mu_0 + [\ln(r/a)]/b 
\end{equation}
at radius $r$, in terms of the spiral scaling constants defined above,
$a$ and $b$. 

The out-of-plane field topology is not well-constrained by previous
observations, but might be expected to have either an even or odd
configuration of the symmetry about the mid-plane \citep[see Fig.~2
  of][]{widrow_2002}. An even configuration corresponds to the case
where the azimuthal (toroidal) component of the field has the same
sign both above and below the mid-plane. The resulting field geometry
has a quadrupole structure in the poloidal field and is the one 
predicted to emerge most naturally from an $\alpha\omega$ dynamo
operating at intermediate to large radii in a galactic disk, where
differential rotation is important \citep{elstner_etal_1992}. An
odd-parity configuration has the opposite signs of the azimuthal field
above and below the mid-plane. The associated structure of the
poloidal field is that of a dipole. This topology may be associated
with the $\alpha^2$ dynamo process that may be dominant within the
circum-nuclear regions where solid body rotation may prevail in
galaxies \citep{elstner_etal_1992}. The $\alpha\omega$ dynamo in
galactic halos may also generate dipolar fields
\citet{sokoloff_shukurov_1990}.

Of stronger relevance is the expected
height above and below the mid-plane at which the polarized
synchrotron emission might originate. From the discussion in
Sect.\,\ref{section:trends}, we expect that polarized emissivity may peak
on either side of the galaxy mid-plane, but that the near-side
component will dominate the detected polarized intensity at GHz
frequencies because of depolarization in the turbulent mid-plane,
  which is an intervenor to the far-side emission. The
Faraday depth distribution will reflect all of the relevant
propagation effects that the emission has experienced. For the
near-side component this will reflect the extended near-side halo of
the target galaxy, while for the far-side component, the additional
dispersion in the mid-plane region will also contribute.

\begin{figure*}
\resizebox{\hsize}{!}{\includegraphics{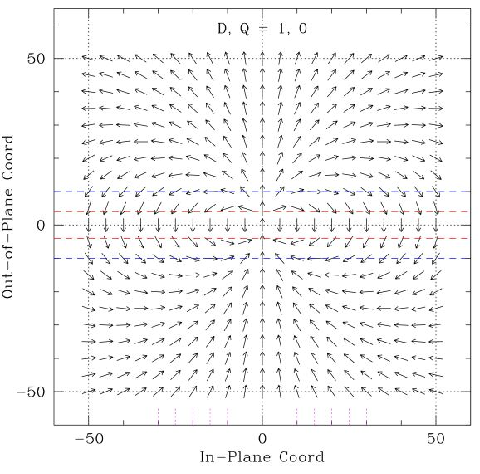},\includegraphics{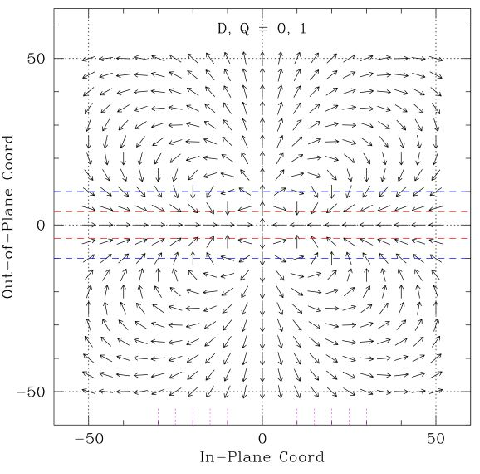},\includegraphics{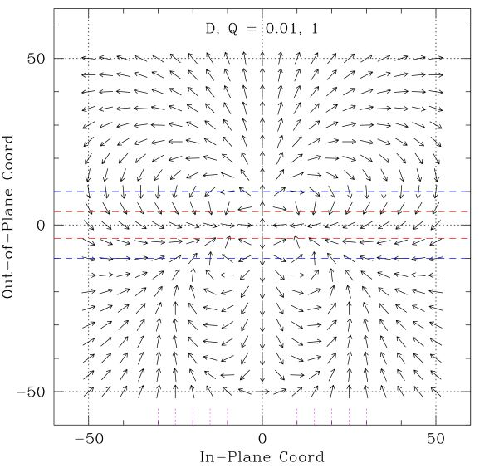}}
\caption{Depiction of the assumed poloidal modification to the out-of-plane field
  topology. A planar logarithmic spiral is modified by the local
  orientation of a dipole or quadrupole field that is symmetric about
  the rotation axis. The panels are labelled with the relative
  strengths of dipole (D) and quadrupole (Q) moments and illustrate a
  pure dipole, pure quadrupole, and a 1:100 mix of dipole and
  quadrupole from left to right. Dashed horizontal lines illustrate
  the heights for which model distributions are shown in
  Fig.~\ref{figure:sims}. }
\label{figure:geom}
\end{figure*}

\begin{figure*}
\resizebox{\hsize}{!}{\includegraphics{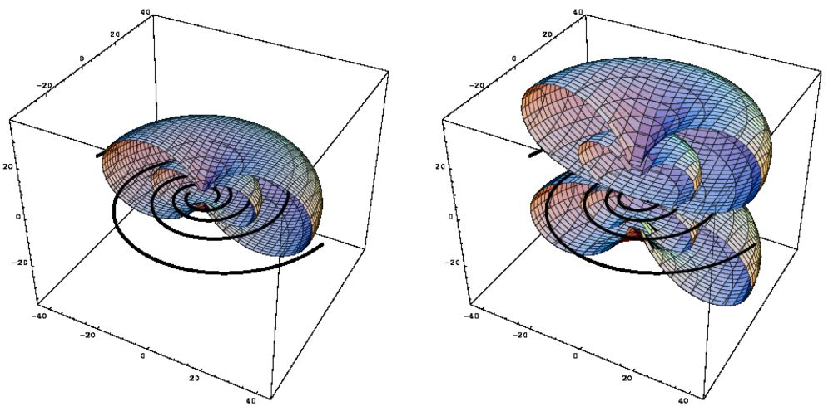}}
\caption{Depiction of the magnetic flux function for a pure dipole
  (left) and quadrupole (right) field. In our modeling we assume that
  a planar logarithmic spiral field, as illustrated with the solid
  lines in the Figure defines the angle $\psi_{xy}$, while the local orientation of
  a dipole or quadrupole field that is symmetric about the rotation
  axis defines the angle, $\psi_z$, out of the plane. }
\label{figure:topol}
\end{figure*}

We extend the usual planar ASS and BSS field topology with the
addition of an out-of-plane component taken from a linear combination
of dipole and quadrupole topologies. The basic equations describing
dipole and quadrupole fields and magnetic flux functions can be found
in \citet{long_etal_2007}. For the simple case under consideration,
there is cylindrical symmetry about the galaxy rotation axis, z. In
terms of the angle from the rotation axis, $\theta$, and the distance
from the origin, $\rho$, the two perpendicular components of the
poloidal magnetic field with a dipole moment, D, and quadrupole
moment, Q, (each with a non-trivial sign) are given by,
\begin{eqnarray}
B_\rho & = & { 2 D cos(\theta) \over \rho^3} + {3 Q [ 3 cos^2(\theta) - 1] \over 4 \rho^4},\\ 
B_\theta & = & {D sin(\theta) \over \rho^3} + {3 Q sin(\theta) cos(\theta) \over 2 \rho^4}.
\label{eqn:dq}
\end{eqnarray}
The corresponding magnetic flux function, $\Psi$, is given by,
\begin{eqnarray}
\Psi & = & {D sin^2(\theta) \over \rho} + {3 Q \over 4 \rho^2} sin^2(\theta) cos(\theta).
\end{eqnarray}
The surfaces defined by $\Psi = constant$ represents example surfaces
on which the magnetic field lines reside. The field components out of-
and within the plane are given by,
\begin{eqnarray}
B_z & = & B_\rho cos(\theta) - B_\theta sin(\theta),\\ 
B_r & = & B_\rho sin(\theta) + B_\theta cos(\theta),
\end{eqnarray}
which yield the total field strength and local orientation angle from,
\begin{eqnarray}
B^2 & = & B^2_z + B^2_r\\ 
\psi_{z} & = & \tan^{-1}(B_z/B_r).
\label{eqn:pz}
\end{eqnarray}
The poloidal field topologies are shown in a plane that includes the
rotation axis in Fig.~\ref{figure:geom} for pure dipole, quadrupole,
and a mixed dipole plus quadrupole configuration. The units of the two
axes in the plot are arbitrary, since the topologies are
self-similar. A three dimensional depiction of the pure dipole and
quadrupole field topologies is given in Fig.~\ref{figure:topol}, where
several surfaces of constant magnetic flux are shown for each
case. 

Combining the planar (Eq.~\ref{eqn:pxy}) and out-of-plane
(Eq.~\ref{eqn:pz}) geometries permits the three right-handed
cartesian components of the modified ASS magnetic field in the frame
of the galaxy to be written as,
\begin{eqnarray}
B_x & = & B\cos(\phi+\psi_{xy})\cos(\psi_z), \\
B_y & = & B\sin(\phi+\psi_{xy})\cos(\psi_z),\\
B_z & = & B\sin(\psi_z).
\label{eqn:ass}
\end{eqnarray}
When viewed at an
inclination, $i$, this yields the observer's frame components,
\begin{eqnarray}
B_{x^\prime} & = & B\cos(\phi+\psi_{xy})\cos(\psi_z), \\
B_{y^\prime} & = & B\sin(\phi+\psi_{xy})\cos(\psi_z)\cos(i)-B\sin(\psi_z)\sin(i), \\
\label{eqn:pass}
B_{z^\prime} & = & B\sin(\phi+\psi_{xy})\cos(\psi_z)\sin(i)+B\sin(\psi_z)\cos(i),
\end{eqnarray}
where the $(x^\prime,y^\prime,z^\prime)$ are the major axis, minor axis, and
line of sight, respectively. If the spiral is a trailing
one (as demonstrated in every kinematically studied galaxy
with a well-defined nearside) then the positive $x^\prime$ axis so defined
corresponds to the receding major axis. The projected parallel and
perpendicular components of the magnetic field and the
orientation angle of $B_{\perp}$ are given by,
\begin{eqnarray}
B_{\parallel} & = & B_{z^\prime} \\
B_{\perp} & = & \sqrt{B^2_{x^\prime}+B^2_{y^\prime}}\\
\chi^\prime_0 & = & \arctan\left(\frac{B_{y^\prime}}{B_{x^\prime}}\right).
\label{eqn:opass}
\end{eqnarray}
We recall that the Faraday depth caused by a magneto-ionic medium is
proportional to $B_{\parallel}$, the polarized intensity is
proportional to $B^{1+\alpha}_{\perp}$, and the intrinsic polarization angle
(giving the E-field orientation) is $\chi_0 = \chi^\prime_0 -
90^\circ$.  

In the corresponding case of a modified BSS magnetic field, we have
\begin{eqnarray}
B_x & = & B\cos(2\phi+\psi_{xy}+\mu)\cos(\psi_z), \\
B_y & = & B\sin(2\phi+\psi_{xy}+\mu)\cos(\psi_z), \\
B_z & = & B\sin(\psi_z).
\label{eqn:bss}
\end{eqnarray}
When viewed at an
inclination, $i$, this yields the observer's frame components,
\begin{eqnarray}
B_{x^\prime} & = & B\cos(2\phi+\psi_{xy}+\mu)\cos(\psi_z), \\
B_{y^\prime} & = & B\sin(2\phi+\psi_{xy}+\mu)\cos(\psi_z)\cos(i)-B\sin(\psi_z)\sin(i), \\
B_{z^\prime} & = & B\sin(2\phi+\psi_{xy}+\mu)\cos(\psi_z)\sin(i)+B\sin(\psi_z)\cos(i),
\end{eqnarray}
and the corresponding $B_{\parallel}$, $B_{\perp}$, and $\chi^\prime_0$
as above.

With these definitions in place, it is possible to explore the
parameter space of modified ASS and BSS field topologies and produce
both images and azimuthal traces of the expected distributions of
$B_{\parallel}$ and $B_{\perp}$. We use these measures as proxies
of the Faraday depth and polarized intensity, respectively.  In the
absence of a model for the spatial distributions of cosmic-ray and
thermal electrons, we do not attempt to reproduce the observables
directly, but merely the systematic modulations
of $P$ and $\Phi$ with azmiuth.

We present images and traces of the ASS and BSS predictions in
Figs.~\ref{figure:psims}--\ref{figure:rhsims}. Each group of $2\times2$
panels in the figure represents a contour plot of $B_{\parallel}$ and
$B_{\perp}$ (at top) and a series of traces of $B_{\parallel}(\phi)$
and $B_{\perp}(\phi)$ at fixed radii (below). The relative weights of
a dipole (D) and quadrupole (Q) field (each including a nontrivial
sign), galaxy inclination, $i$, spiral pitch angle, $\psi_{xy}$
(positive for counter-clockwise and negative for clockwise), and
distance(s), Z or (Z1 and Z2), from the mid-plane (positive toward the
observer and negative away) are indicated at the top of the contour
plots, together with an ASS or BSS designation. Two representative
spirals are drawn for reference, both directed inwardly for the ASS plots
and one inward and the other outward for the BSS plots.  The inclination is
defined such that for kinematically trailing spirals, the receding and
approaching ends of the major axis are as indicated. The ``near'' and
``far'' designations in the plots also aid in demonstrating the
spatial orientation of the disk unambiguously. An ellipse that is
offset from the disk by the distance Z is drawn to demonstrate its
location relative to the mid-plane. The radii at which the azimuth
plots were made are marked by the same linetype on the contour
plots. The azimuth angle increases counter-clockwise from the receding
major axis.

We also plot the projected angle of $B_{\perp}$,
$\chi^\prime_0(\phi)$, within the same panel that presents the trace
of $B_{\perp}(\phi)$ using the right-hand scale. The
$\chi^\prime_0(\phi)$ traces are only plotted for the ASS cases, since
the BSS cases vary so dramatically with radius.

For simplicity, we begin with a fixed spiral pitch angle,
$\psi^\prime_{xy}~=~20^\circ$, since this corresponds well with the
average measured value for our sample galaxies
\citep[cf.][]{kennicutt_1981}, which vary from about 15--25$^\circ$,
and we begin by considering counter-clockwise (CCW) spirals. Other
values will also be considered below. First we 
consider the predictions for the simple planar ASS and BSS
spirals in Fig~\ref{figure:psims}. A radial spiral scaling constant, $a~=~5$,
and a galaxy disk radius, $r_{max}~=~35$, are assumed for
illustration. These choices have no effect on the results. 

\subsection{Planar models\label{subsection:planarmod}}

Planar, axisymmetric field topologies yield projected field
components (as shown in the left-hand groups of
Fig.~\ref{figure:psims}) that are independent of radius and have very
simple symmetries, $B_{\parallel}(\phi)$ having positive and negative
excursions that are fully symmetric about zero. The single positive
peak occurs at $\phi~\sim~\psi^\prime_{xy}$ offset from the
approaching major axis, the negative peak being opposite to this
\citep[as also demonstrated by][]{krause_1990}. The $B_{\perp}(\phi)$
component has
a complimentary behavior with two equal minima slightly offset from
both major axes, and two equal maxima slightly offset from both minor
axes. As the inclination increases from face-on toward edge-on, the
amplitude of the $B_{\parallel}(\phi)$ and $B_{\perp}(\phi)$
modulation increases, although there is no change in the azimuthal
location of either the maxima or minima. The projected orientation of the
plane-of-sky field, $\chi^\prime_0(\phi)$, has the expected linear
variation with azimuth for a nearly face-on geometry, which becomes
increasingly non-linear as the inclination increases.

Changing the sense of the field to be outwardly directed, rather than
inwardly directed, changes only the sign of $B_{\parallel}(\phi)$, and
leaves $B_{\perp}(\phi)$ unchanged. This is true for all of the
field geometries we consider.

The basic BSS projected field patterns and their variation with galaxy
inclination are illustrated in the right-hand groups of
Fig.~\ref{figure:psims}. Because of the modulation of field sense with
azimuth along the family of spirals, there is a strong radial
dependence of the projected field components and their variation with
azimuth. $B_{\parallel}(\phi)$ has positive and negative excursions
that are still fully symmetric about zero, but it exhibits two
equal maxima and two equal minima (rather than only one for ASS),
which are located at different azimuth depending on the
radius. $B_{\perp}(\phi)$ displays four equal maxima at the
azimuthal angles where $B_{\parallel}(\phi)$ has its zero crossings,
and four equal minima in-between (at the locations of maximum and
minimum $B_{\parallel}$). As with the basic ASS fields, a higher
inclination yields an increased fractional modulation of
$B_{\parallel}(\phi)$ and $B_{\perp}(\phi)$, while not affecting
the location of those excursions. The strong radial dependence of
these azimuthal patterns implies that if there were significant
averaging of radii (by even as little as 20\% in $\Delta r/r$ for a
pitch angle of $\psi^\prime_{xy}~\sim~20^\circ$) then most of the
predicted modulation would disappear. This is particularly true of
$B_{\perp}(\phi)$, for which the predicted modulation is both
intrinsically smaller and twice as rapid as that of
$B_{\parallel}(\phi)$. The strong radial dependance of BSS models
severely limits their predictive power for the general patterns of
azimuthal modulation that are observed in galaxies and so will not be
considered further.

The azimuthal modulation of $B_{\parallel}$ and $B_{\perp}$ in the
planar ASS and BSS spiral topologies is apparently inadequate for
describing the general patterns noted above in
Sect.\,\ref{section:trends}. Neither of these topologies predict a clear
distinction between the receding and approaching kinematic major axis
in terms of either the intrinsic brightness of polarized intensity
or the Faraday depth that it may encounter.

\begin{figure*}
\resizebox{\hsize}{!}{\includegraphics{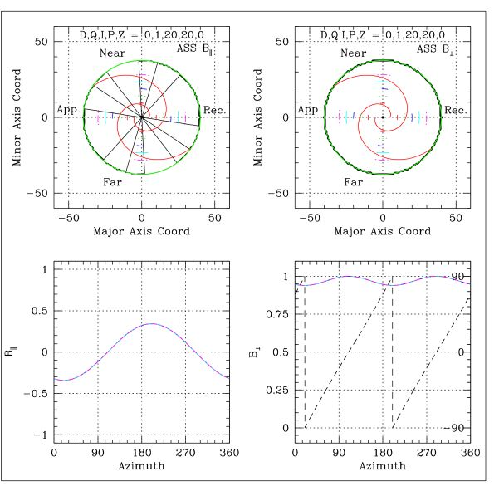},\includegraphics{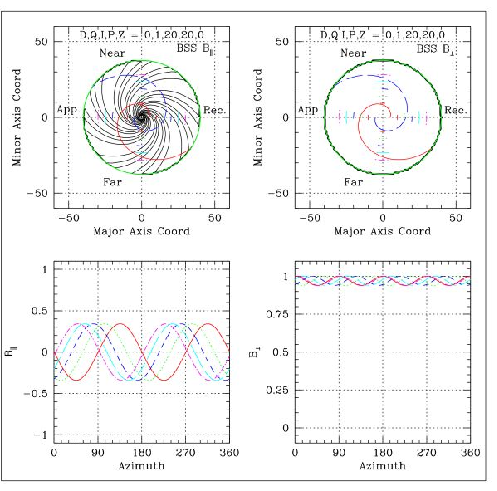}}
\resizebox{\hsize}{!}{\includegraphics{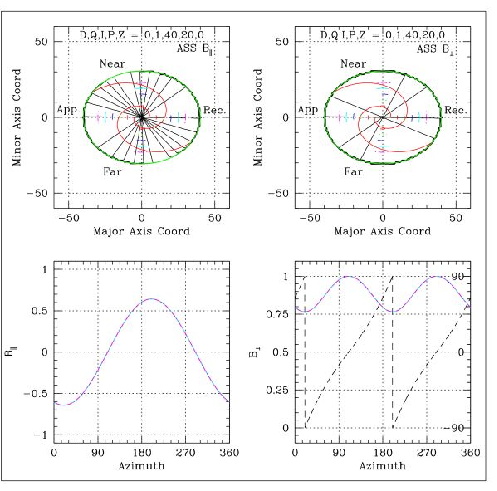},\includegraphics{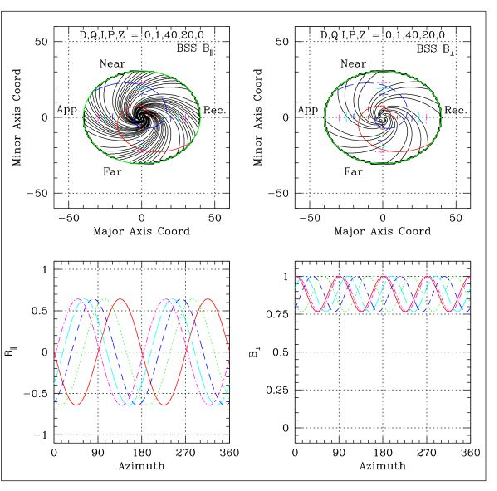}}
\caption{Simulated field distributions. Each group of $2\times2$
  panels presents contour plots (top) and azimuthal traces (bottom) of
  $B_\parallel$ (left) and $B_\perp$ (right). The dipole and
  quadrupole moments (D, Q), galaxy inclination (I), spiral pitch
  angle (P), and height above the galaxy mid-plane (Z) are indicated at
  the top of each contour plot. Contours are drawn at $\pm$10,
  $\pm$20, $\pm$30, $\dots$ $\pm$90\% of $|B|$. Two reference spirals
  are drawn, both with field directed inward ($B~=~+1$) for the
  axisymmetric (ASS) case and one in and the other out for the
  bisymmetric (BSS) case. The approaching and receding major axis of
  the galaxy are indicated for trailing spirals. Azimuth is measured
  CCW from the receding major axis. Azimuth traces are drawn for the
  radii indicated in the contour plots. A trace of the orientation of
  $B_\perp$, $\chi^\prime_0(\phi)$, is given in the $B_\perp$ panel
  for ASS models with a dashed linetype using the right-hand scale in
  degrees. Here we compare the ASS (left hand groups) and BSS (right hand
  groups) planar models (Z~=~0). The galaxy inclination is 20$^\circ$
  in the upper series of groups and 40$^\circ$ in the lower
  series. The spiral pitch angle is 20$^\circ$.}
\label{figure:psims}
\end{figure*}

\addtocounter{figure}{-1}
\begin{figure*}
\resizebox{\hsize}{!}{\includegraphics{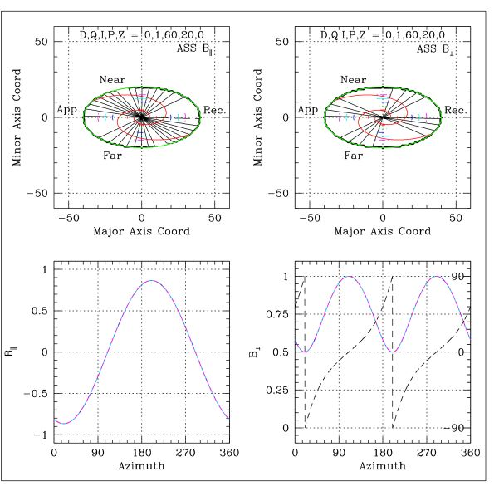},\includegraphics{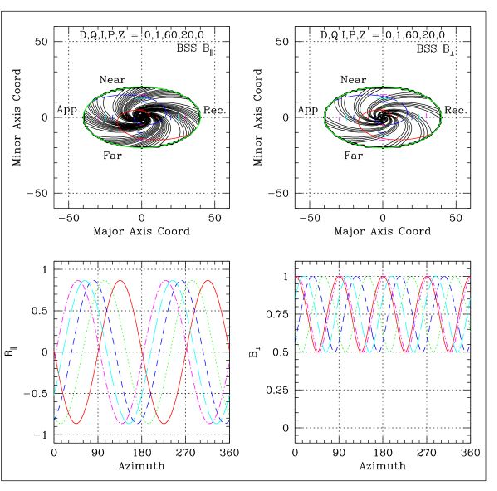}}
\resizebox{\hsize}{!}{\includegraphics{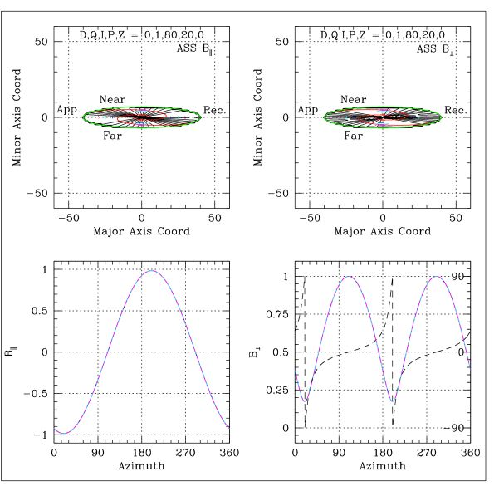},\includegraphics{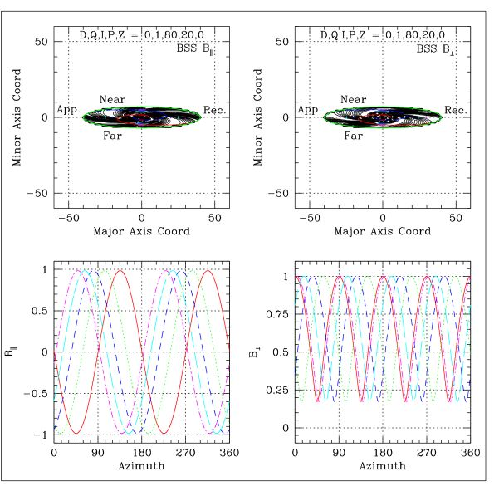}}
\caption{(continued) Simulated field distributions. Here we compare
  ASS (left hand groups) and BSS (right hand groups) planar models
  (Z~=~0). The galaxy inclination is 60$^\circ$ in the upper series of
  groups and 80$^\circ$ in the lower series. }
\end{figure*}

\subsection{Thick disk models\label{subsection:thickmod}}

In Figs.~\ref{figure:sims} -- \ref{figure:rhsims}, we explore the
extension of the basic ASS topology with either a dipole (D~=~1,
left-hand groups) or quadrupole (Q~=~1, right-hand groups)
modification to the out-of-plane field, as given by
Eq.~\ref{eqn:ass}. In keeping with the discussion of
Sect.\,\ref{section:trends}, we calculate a projected distribution that
is populated within a specified range of mid-plane heights extending
from Z1 to Z2 on the near-side of the galaxy and from $-$Z2 to $-$Z1
on the far-side. For lines-of-sight that intersect the mid-plane disk
within the nominal galaxy radius, only the near-side region (Z~$>$~0)
contributes to the integral, to approximate the effect of mid-plane
depolarization. For other lines-of-sight, out to 1.2 times the nominal
galaxy radius, both the nearside and farside zones are
integrated. The integral is determined by first evaluating
$B_{\parallel}$ and $B_{\perp}$ at a sequence of finely-sampled
line-of-sight depths and then forming the average. The two lower
panels of each group were modified to allow more direct
comparison with the data of Fig.~\ref{figure:fdpofaz} by plotting the
mean modeled quantity within azimuthal wedges rather than the
azimuthal traces at discrete radii. The error bars represent the RMS
variation within each wedge, which are principally caused by a dependence
on radius.

We first consider the dipole and quadrupole models that are
integrated over mid-plane heights of $|z|~=~2 \rightarrow 4$ (about 10\% of the
disk radius) in Fig.~\ref{figure:sims}.  Although the azimuthal trace
of projected field modulation with azimuth show some variation in
amplitude with radius (as indicated by the error bars), there are
well-defined maxima and minima. The local modification to the planar
spiral field is illustrated in Fig.~\ref{figure:geom} by the vector
orientations along the horizontal line drawn at Z~=~$+$4. The addition
of a poloidal field component has introduced several notable
differences from the planar case. Although $B_{\parallel}(\phi)$
exhibits maxima and minima at the same azimuth as previously (near the
major axes), it no longer has excursions that are symmetric about
$B_{\parallel}~=~0$, $B_{\parallel}(\phi)$ being offset to negative
values of $B_{\parallel}$ (for an inwardly-directed spiral field) and
showing complementary modulation in the dipole and quadrupole cases. The
character of $B_{\perp}(\phi)$ is more substantially modified by the
addition of a poloidal field. We first consider the quadrupole
case (Q~=~1) shown in the right-hand groups of Fig.~\ref{figure:sims}.

At low inclinations, a single minimum in $B_{\perp}$ occurs
near the receding major axis at $\phi~\sim~\psi^\prime_{xy}$ and a
broad single peak centered near the approaching major axis. As the
inclination increases, the single global minimum in $B_{\perp}$
remains close to the receding major axis, while the single broad maximum
divides into two maxima that separate and shift toward the two ends
of the minor axis. This is exactly the trend shown by the data in
Fig.~\ref{figure:fdpofaz}.  
The other noteworthy attribute of these distributions is that the
largest excursion in $|B_{\parallel}|$ is found on the receding major
axis at all inclinations, representing the greatest Faraday depth at this
azimuth for coextensive emitting and dispersive media. For low
inclinations, there is a broad minimum in $|B_{\parallel}|$ at the
approaching major axis, which becomes a secondary peak at this
location (of opposite sign) as the inclination increases. This
roughly agrees with the patterns seen for the majority
of lower inclination galaxies in Fig.~\ref{figure:fdpofaz}, these
models still being more symmetric than the data, which exhibit a much
clearer nearside/farside asymmetry. 

For the dipole case (D~=~1), shown in the left-hand groups of
Fig.~\ref{figure:sims}, the receding and approaching major axes are
reversed, since the poloidal field component of the dipole at positive
mid-plane offsets is directed outward and not inward (as shown in
Fig.~\ref{figure:geom}). For this case, it is the {\it approaching\ }
major axis that is predicted to have a global minimum in $B_{\perp}$,
and hence also the minimum intrinsic polarized intensity. In a similar way, it is the
{\it approaching\ } major axis for which the greatest value
of $|B_{\parallel}|$, hence the largest associated Faraday
depth is found. The degree of modulation with azimuth is rather modest for the
dipole case integrated over these heights.

From this comparison, it is clear that a large-scale quadrupole ASS
field topology provides an excellent model for explaining the
modulation of polarized intensity with azimuth, while the dipole
clearly does not. The modulation of Faraday depth with azimuth is also
explained qualitatively for the low inclination systems, but is not
reproduced in detail. The crucial attributes of the quadrupole model
in matching the observed azimuthal modulation pattern, $<P(\phi)>$,
are the magnitude and particularly the sign of of the angle $\psi_z$
as given by Eq.~\ref{eqn:pz}. For typical values of the spiral pitch
angle, $|\psi^\prime_{xy}| < 25^\circ$, the projected field component,
$B_{y^\prime}$ of Eq.~\ref{eqn:pass} has minima and maxima near $\phi~=~0$
and $180^\circ$. The ``hour-glass'' shape of the quadrupole field yields
a positive sign of $\psi_z$ for small positive values of $z$,
i.e. toward the observer, and this is what yields a minimum of
$B_{\perp}$ near $\phi~=~0$ for the nearside emission. The ``donut''
shape of the dipole field, on the other hand, yields a negative sign of
$\psi_z$ for small positive values of $z$, resulting in a maximum of
$B_{\perp}$ near $\phi~=~0$.

\begin{figure*}
\resizebox{\hsize}{!}{\includegraphics{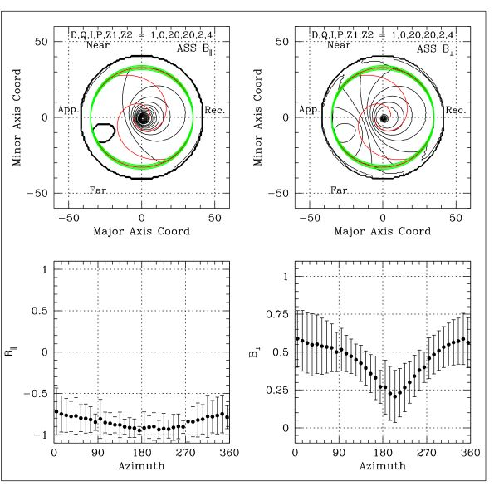},\includegraphics{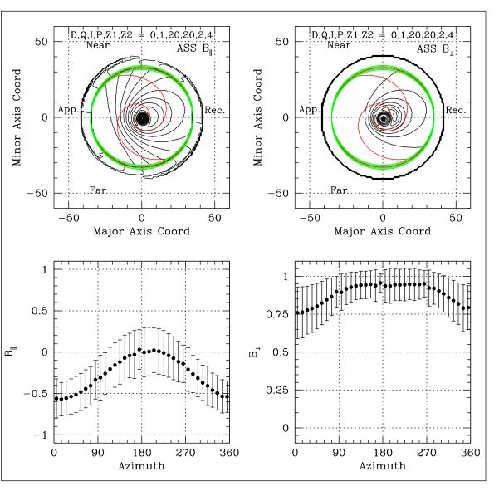}}
\resizebox{\hsize}{!}{\includegraphics{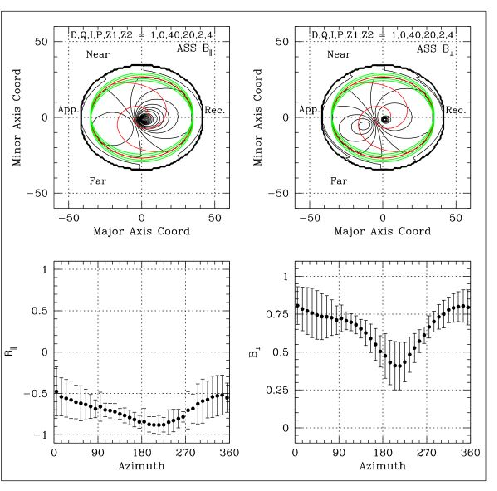},\includegraphics{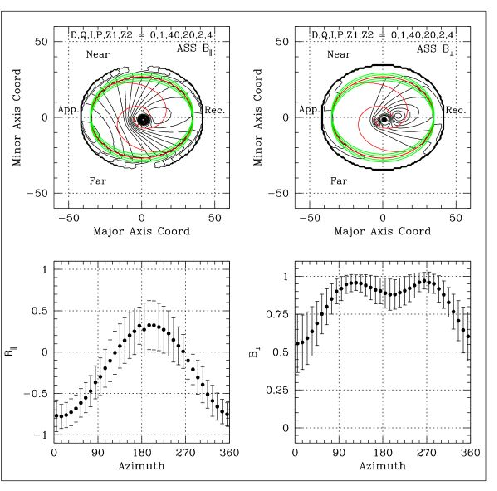}}
\caption{Simulated field distributions (as in
  Fig.~\ref{figure:psims}). Here we compare dipole- (left hand
  groups) and quadrupole- (right hand groups) ASS models integrated
  over a zone from Z1 to Z2 ($|$Z$|$~=~2--4). For those lines-of-sight
  that intersect the mid-plane of the disk, only positive values of Z
  are included in the integral. The galaxy inclination is 20$^\circ$
  in the upper series of groups and 40$^\circ$ in the lower series. }
\label{figure:sims}
\end{figure*}

\addtocounter{figure}{-1}
\begin{figure*}
\resizebox{\hsize}{!}{\includegraphics{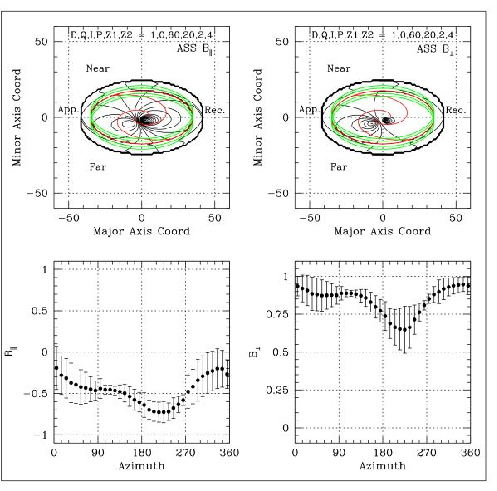},\includegraphics{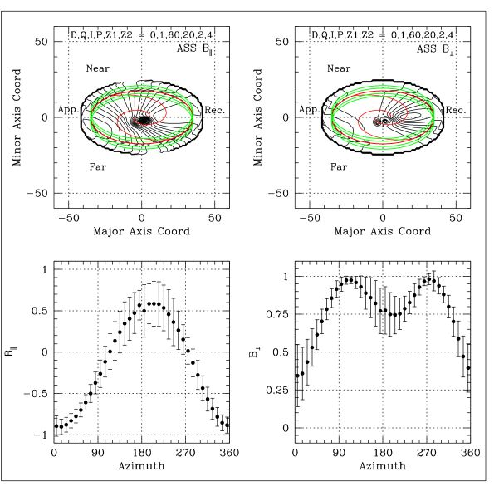}}
\resizebox{\hsize}{!}{\includegraphics{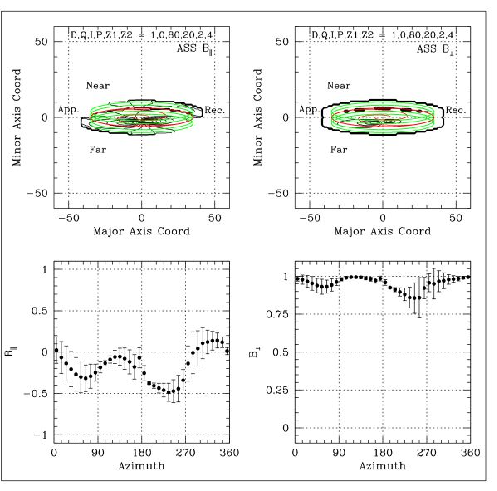},\includegraphics{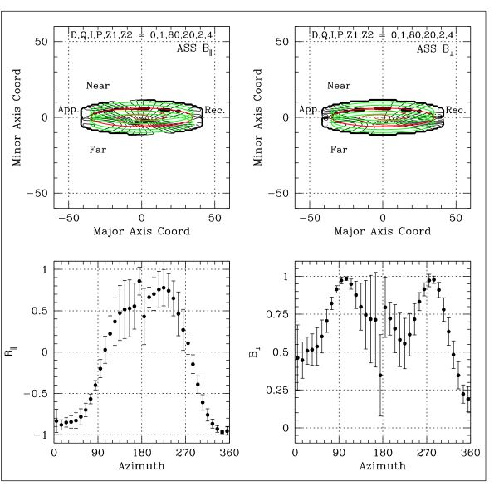}}
\caption{(continued) Simulated field distributions. The galaxy inclination
  is 60$^\circ$ in the upper series of groups and 80$^\circ$ in the
  lower series.}
\end{figure*}

\subsection{Halo models\label{subsection:halomod}}

We now consider dipole and quadrupole models that extend to more
substantial mid-plane heights in an attempt to reproduce the azimuthal
modulations of Faraday depth seen in the data. In
Fig.~\ref{figure:hsims}, we illustrate the result of integrating the
same models just considered over mid-plane heights of $|z|~=~4
\rightarrow 10$ (or about 30\% of the disk radius). By considering
first the quadrupole models in the right-hand panels, we see
substantial similarity between these models and those originating
closer to the mid-plane (Fig.~\ref{figure:sims}). The modulation
pattern in $B_{\perp}(\phi)$ has a slightly smaller amplitude. The
modulation pattern of the halo $B_{\parallel}(\phi)$, while similar to
its thick disk counterpart, has a clear asymmetry in the approaching
major axis minimum about an azimuth of 180$^\circ$, which is
reminiscent of what is seen in the data of
Fig.~\ref{figure:fdpofaz}. The pattern shown is for a positive pitch
angle of 20$^\circ$ (CCW spiral), while for a negative pitch angle (CW
spiral) it is mirrored about an azimuth of 180$^\circ$. We recall that
changing the sign of the spiral field from inward directed (as shown)
to outward directed changes only the sign of
$B_{\parallel}(\phi)$. Qualitative agreement of the halo predictions
for the quadrupole with the basic $<\Phi(\phi)>$ patterns is
reasonable for many of the low inclination galaxies of our sample. The
dipole halo models (shown in the left-hand panels) with their maximal
excursion near an azimuth of 180$^\circ$ reproduces the observations
less successfully.

Although the models considered to this point can reproduce the 
general pattern of $<P(\phi)>$ modulation at all galaxy inclinations
and the $<\Phi(\phi)>$ modulations for many low inclination targets,
they have not provided any agreement with the distinctive doubly
peaked pattern of $<\Phi(\phi)>$ around the minor axes seen in our
four highest inclination galaxies. We now consider an additional variant
of our halo models in an attempt to reproduce those $<\Phi(\phi)>$
patterns. In Fig.~\ref{figure:rhsims} we consider dipole and
quadrupole models for an extremely large pitch angle of 85$^\circ$, to
illustrate the impact of a planar field component that is essentially
radial. The distinctive doubly-peaked pattern of $<\Phi(\phi)>$ at
large inclinations can be reasonably reproduced in this case, but only
for the dipole topology.

\begin{figure*}
\resizebox{\hsize}{!}{\includegraphics{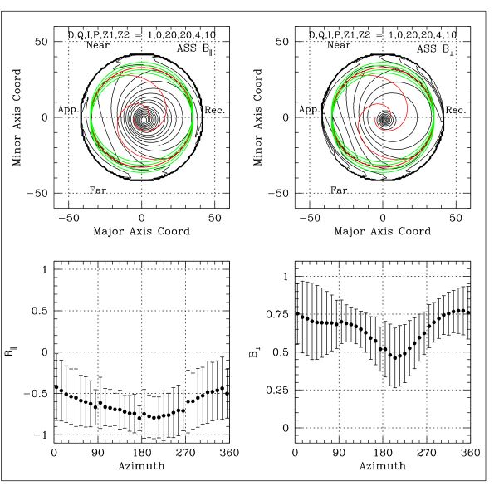},\includegraphics{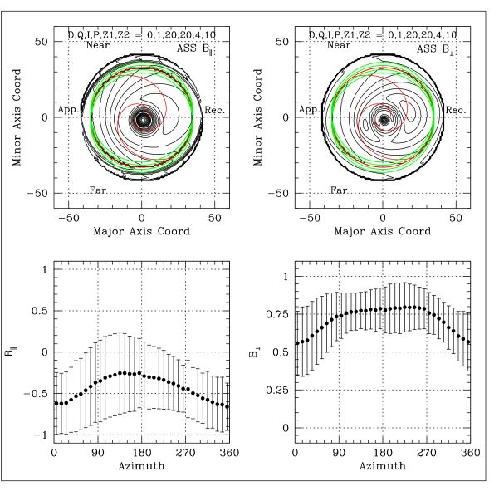}}
\resizebox{\hsize}{!}{\includegraphics{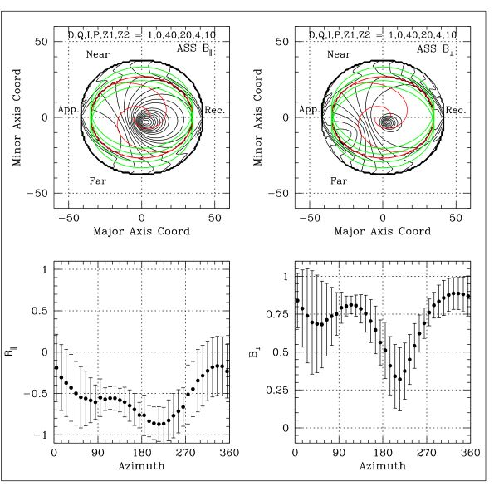},\includegraphics{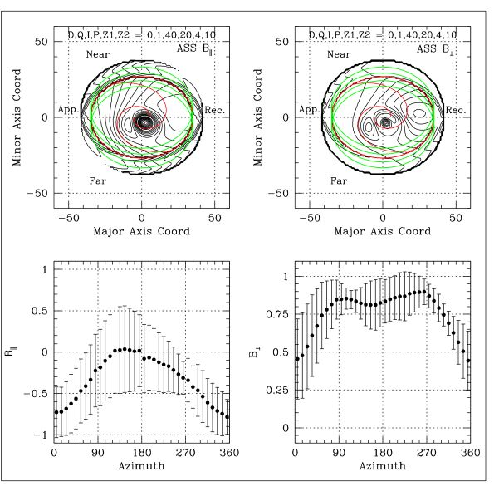}}
\caption{Simulated field distributions (as in
  Fig.~\ref{figure:psims}). Here we compare dipole- (left hand
  groups) and quadrupole- (right hand groups) ASS models integrated
  over a zone from Z1 to Z2 ($|$Z$|$~=~4--10). For those
  lines-of-sight that intersect the mid-plane of the disk, only
  positive values of Z are included in the integral. The galaxy
  inclination is 20$^\circ$ in the upper series of groups and
  40$^\circ$ in the lower series. }
\label{figure:hsims}
\end{figure*}

\addtocounter{figure}{-1}
\begin{figure*}
\resizebox{\hsize}{!}{\includegraphics{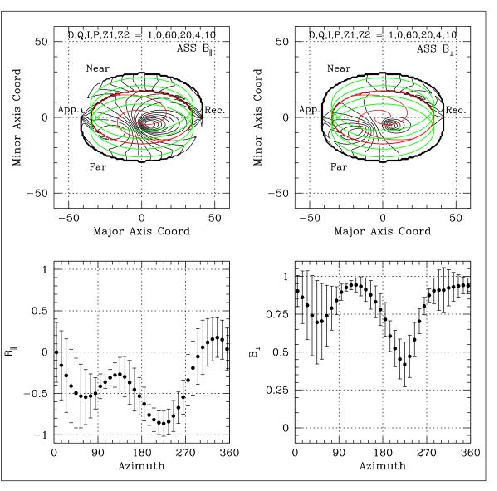},\includegraphics{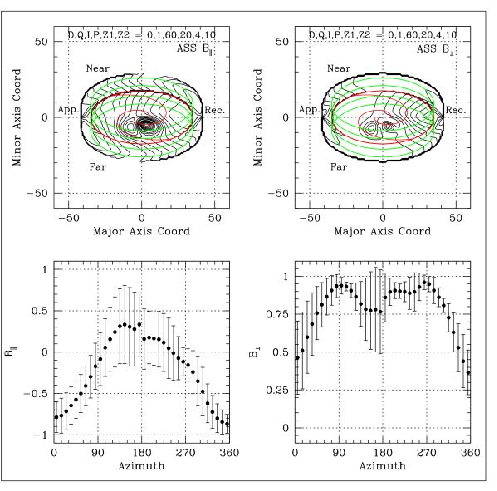}}
\resizebox{\hsize}{!}{\includegraphics{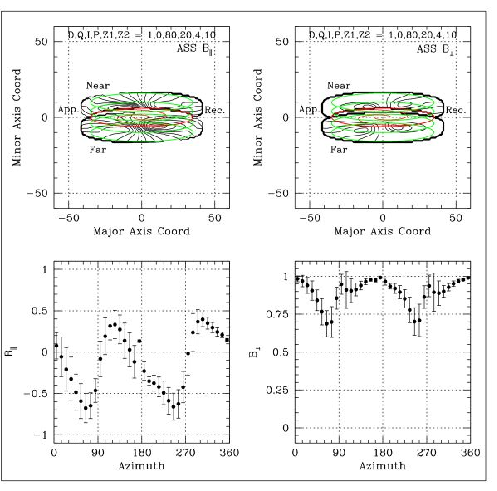},\includegraphics{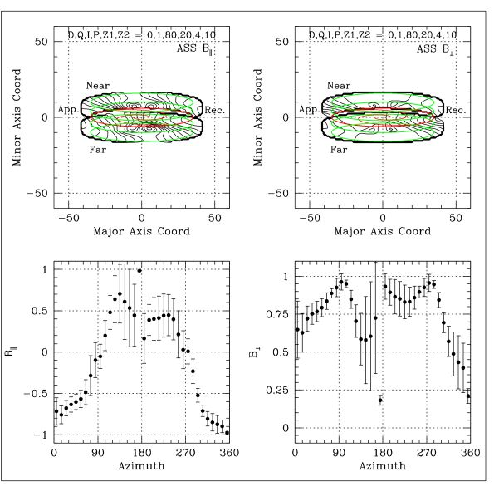}}
\caption{(continued) Simulated field distributions. The galaxy inclination
  is 60$^\circ$ in the upper series of groups and 80$^\circ$ in the
  lower series.}
\end{figure*}

\begin{figure*}
\resizebox{\hsize}{!}{\includegraphics{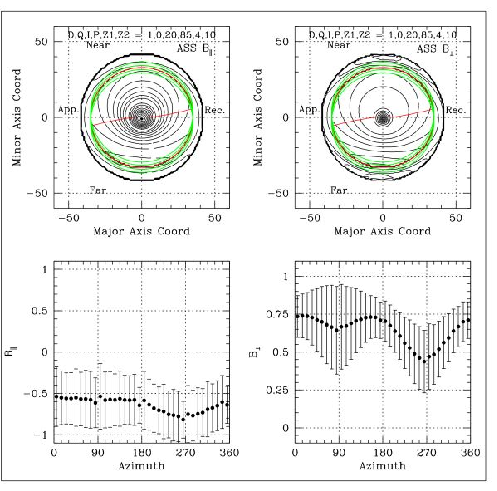},\includegraphics{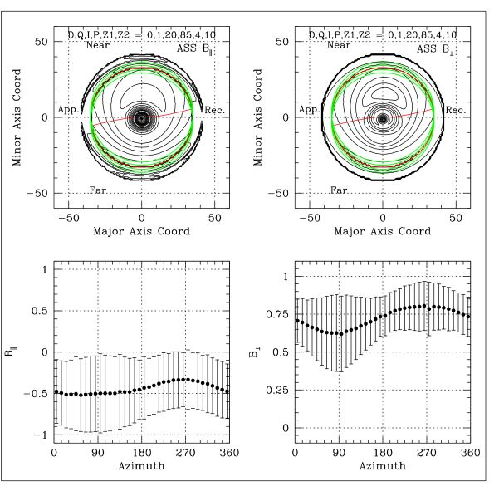}}
\resizebox{\hsize}{!}{\includegraphics{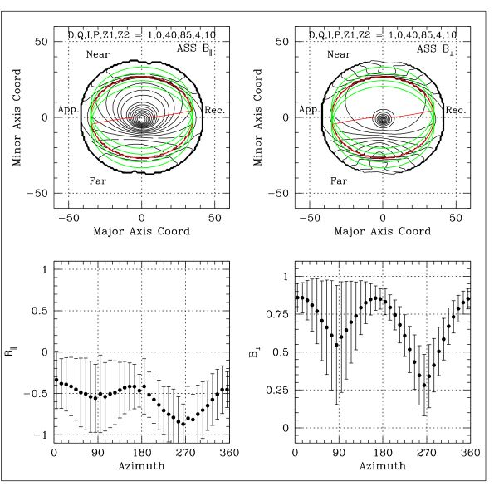},\includegraphics{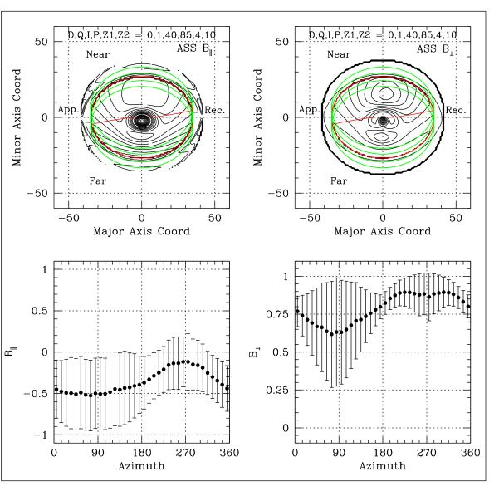}}
\caption{Simulated field distributions (as in
  Fig.~\ref{figure:psims}). Here we compare dipole- (left hand
  groups) and quadrupole- (right hand groups) ASS models integrated
  over a zone from Z1 to Z2 ($|$Z$|$~=~4--10) with a large pitch angle
  (P~=~85$^\circ$). For those lines-of-sight that intersect the
  mid-plane of the disk, only positive values of Z are included in the
  integral. The galaxy inclination is 20$^\circ$ in the upper series
  of groups and 40$^\circ$ in the lower series. }
\label{figure:rhsims}
\end{figure*}

\addtocounter{figure}{-1}
\begin{figure*}
\resizebox{\hsize}{!}{\includegraphics{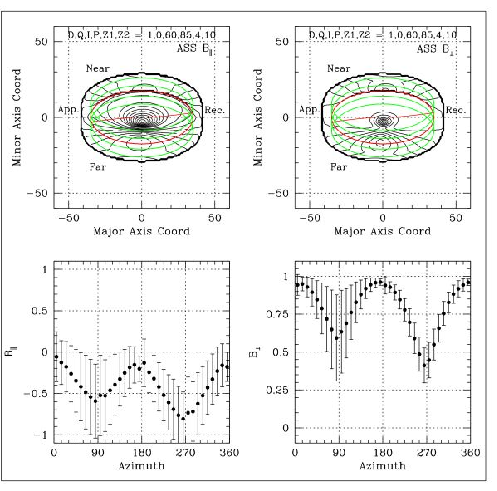},\includegraphics{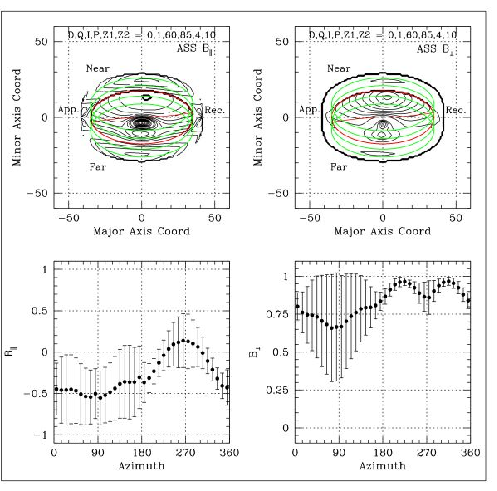}}
\resizebox{\hsize}{!}{\includegraphics{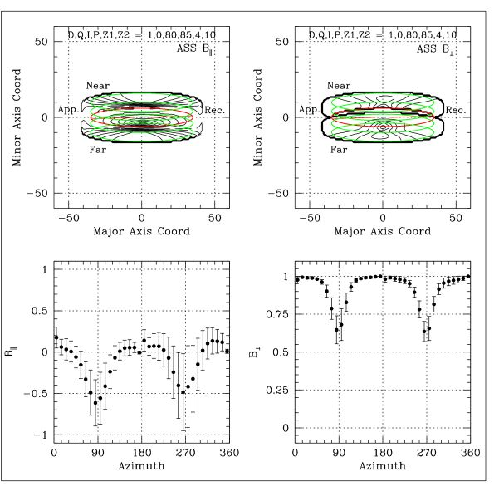},\includegraphics{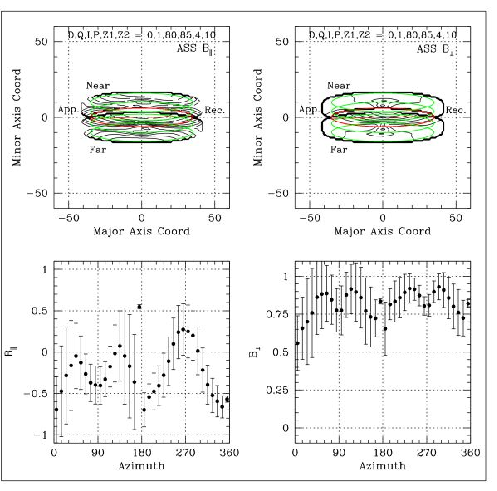}}
\caption{(continued) Simulated field distributions. The galaxy inclination
  is 60$^\circ$ in the upper series of groups and 80$^\circ$ in the
  lower series.}
\end{figure*}

\section{Discussion\label{section:disc}}

Our analysis of the projected three-dimensional magnetic field topologies
presented in Sect.~\ref{section:Bdist} and their predicted observable consequences
for the azimuthal modulation, $B_{\parallel}(\phi)$ and
$B_{\perp}(\phi)$, has provided a plausible explanation of the very
general observed trends noted in Sect.~\ref{section:trends}. A
self-consistent scenario has emerged that accounts for the polarized
intensity and its Faraday dispersion observed at GHz frequencies from
galaxy disks. The detected polarized intensity is dominated by a zone
of emissivity above the mid-plane on the side of the galaxy facing the
observer (at a height of perhaps 5 -- 10 \% of the disk radius). This
thick disk emission arises in a region that is dominated by an
axisymmetric spiral with an out-of-disk quadrupole topology, which is
responsible for a distinctive modulation of $B_{\perp}(\phi)$ and its
variation with galaxy inclination. This emission is affected by only a modest
amount of Faraday dispersion, of a few tens of rad~m$^{-2}$, within
the nearside halo of the galaxy in its subsequent propagation. For
the majority of low to modest inclination galaxies ($\le60^\circ$),
the dispersive foreground topology is consistent with an extension of
the thick disk ASS quadrupole out to larger heights above the
mid-plane (of perhaps 30\% of the disk radius). 

The most highly inclined galaxies of our sample require an alternative
halo field topology, in the form of a radially-dominated
dipole, which yields a distinctive doubly-peaked modulation of
$\Phi(\phi)$. It seems significant that in many or possibly all of the
highly inclined galaxies of our sample there is evidence of a
significant circum-nuclear outflow component to the polarized emission, in
addition to that of the disk. This circum-nuclear component would
quite naturally be expected to be associated with a dipole, rather
than a quadrupole field, in view of the likely dominance of the
$\alpha^2$ over the $\alpha\omega$ dynamo mechanism at small galactic
radii \citep[e.g.][]{elstner_etal_1992}. This circum-nuclear dipole
field would also be less likely to have any association with the
spiral pitch angle of the disk given its origin. Because of the
shallower roll-off with radius of a dipole compared to the quadrupole
field (see Eq.~\ref{eqn:dq}), a dipole component may
come to dominate the halo field of the associated galaxy when both are
present. \citet{sokoloff_shukurov_1990} also argued that a
$\alpha\omega$ dynamo operating in the halo would directly
produce a dipole field. Non-stationary global halo models
\citep[e.g.][]{brandenburg_etal_1992} may also provide a natural
explanation of the dipole signature on the largest scales.

In addition to the bright polarized emission originating in the
nearside, the corresponding rear-facing region of polarized
emissivity of the thick disk can also be detected in relatively
face-on galaxies if sufficient sensitivity is available.  This
emission is substantially weaker, by a factor of 4 -- 5, and consistent
with depolarization caused by fluctuations in the magneto-ionic medium of
the mid-plane on scales smaller than a pc. This fainter polarized
component is affected by much greater Faraday dispersion, corresponding to
both plus and minus 150 -- 200~rad~m$^{-2}$ in its propagation
through the dense mid-plane plasma, as well as the near-side
halo. These two maxima (one positive and one negative) in Faraday
depth are aligned approximately with the major axes of each galaxy,
and have approximately symmetric excursions about the Galactic
foreground value. This pattern is consistent with the 
expectation for a simple planar ASS field in the galaxy disk.

Future observations of nearby galaxy disks at frequencies below 200
MHz, such as with the upcoming LOFAR facility, will likely detect
net polarized emission from even larger heights above the galaxy
mid-plane and exclusively from regions unobstructed by the mid-plane
in projection. A good indication for the predicted observables is
given in Fig.~\ref{figure:hsims} in which we present model
integrations of the upper halo (out to 30\% of the disk radius).

\begin{acknowledgements}
The Westerbork Synthesis Radio Telescope is operated by ASTRON (The
Netherlands Institute for Radio Astronomy) with support from the
Netherlands Foundation for Scientific Research (NWO).
\end{acknowledgements}

\bibliographystyle{aa}
\bibliography{singsintrp}

\end{document}